\documentclass[conference]{IEEEtran}
\IEEEoverridecommandlockouts

\usepackage{graphicx}
\usepackage{balance}
\usepackage{amsmath}
\usepackage{amssymb}
\usepackage{multicol}

\usepackage{bibunits}
\usepackage{hyperref}
\usepackage{epsf}
\graphicspath{{./figures/}}

\usepackage[usenames]{color}

\begin{document}

\title{ Analysis and Optimization of a Frequency-Hopping Ad Hoc Network in Rayleigh Fading }
 \author{\authorblockN{Salvatore Talarico and Matthew C. Valenti }
\authorblockA{West Virginia University\\
Morgantown, WV\\
stalari1@mix.wvu.edu,mvalenti@csee.wvu.edu}
\and\authorblockN{Don Torrieri}
\authorblockA{U.S. Army Research Laboratory\\
Adelphi, MD\\
dtorr@arl.army.mil}
\thanks{This work was supported in part by the National Science Foundation under Award No. CNS-0750821 and by the United States Army Research Laboratory under Contract W911NF-10-0109. }

}

\maketitle

\begin{abstract}
This paper proposes a new method for optimizing frequency-hopping ad hoc networks in the presence of Rayleigh fading.  It is assumed that the system uses a capacity-approaching code (e.g., turbo or LDPC) and noncoherent binary continuous-phase frequency-shift keying (CPFSK) modulation.  By using transmission capacity as the performance metric, the number of hopping channels, CPFSK modulation index, and code rate are jointly optimized.   Mobiles in the network are assumed to be uniformly located within a finite area.  Closed-form expressions for outage probability are given for a network characterized by a physical interference channel. The outage probability is first found conditioned on the locations of the mobiles, and then averaged over the spatial distribution of the mobiles. The transmission capacity, which is a measure of the spatial spectral efficiency, is obtained from the outage probability.  The transmission capacity is modified to account for the constraints of the CPFSK modulation and capacity-approaching coding.  Two optimization methods are proposed for maximizing the transmission capacity.  The first is a brute-force method and the second is a gradient-search algorithm.  The results obtained from the optimization shed  new insight into the fundamental tradeoffs among the number of frequency-hopping channels, the modulation index, and the rate of the error-correcting code.
\end{abstract}

\section{Introduction} \label{Section:Intro}
Ad hoc networks comprise mobiles that communicate without centralized control or a pre-existing infrastructure.  The preferred channel access for ad hoc networks is direct-sequence or frequency-hopping (FH) spread spectrum.  This paper focuses specifically on frequency-hopping spread spectrum ad hoc networks.   Such networks are characterized by independent, identical, FH radios that share the same carriers and frequency channels, and are nearly stationary in location over a single hop duration.

The first part of this paper is concerned with the analysis of the outage probability of FH networks, where outage probability is the probability that the signal-to-noise-and-interference ratio (SINR) falls below a predetermined threshold.  By limiting the fading to be slow Rayleigh fading and excluding shadowing, the paper first presents an exact closed-form expression for the outage probability conditioned on the locations of the interferers.


The interferers are assumed to be uniformly distributed in an annular area, where the inner radius is a minimum interferer  distance that could be imposed by an interference-avoidance protocol \cite{hasan:2007}, such as carrier-sense multiple access, and the outer radius is the maximum distance set by the network's geographic footprint.   By averaging over the uniform locations of the interferers, the spatially averaged outage probability is obtained in closed form.    A distinguishing feature of this paper is that it considers networks of limited area, in contrast with the current popular literature, which typically assumes networks of infinite extent (e.g., \cite{andrews:2010}, \cite{win:2009}).

The number of mobiles in the network may be either fixed or random.  Initially, a fixed number of mobiles is assumed, in which case the mobile locations are a realization of a binomial point process (BPP).  Next, it is assumed that the number of mobiles is Poisson distributed, in which case the mobile locations are a realization of a Poisson point process (PPP).   Considering a PPP allows us to obtain results that are consistent with the current popular literature, and in fact, our results coincide with the infinite-network results of \cite{baccelli:2006, Linnartz:1992, Zorzi:1995} when we let the network boundary extend to infinity.  However, the BPP results are of practical interest because of limitations to the PPP model.  The most significant limitation to the PPP model is that it allows an unbounded number of users, which is not possible in a finite network.  However, the PPP has been favored in the literature because it enables the use of Campbell's theorem \cite{stoyan:1996}, which often leads to tractable mathematical expressions that vastly simplify the performance analysis.

Having found the spatially averaged outage probability under the BPP and PPP models, the paper next derives closed-form expressions for the {\em transmission capacity} \cite{weber:2005}, which is the spatial spectral efficiency; i.e., the rate of successful transmissions per Hz and $m^2$.  We propose a modification to the transmission capacity metric of \cite{weber:2005} that accounts for modulation and coding constraints.  The utility of the modulation-constrained transmission capacity is that it can be used to optimize the main parameters that influence the network's performance.  The network's performance depends on several parameters related to the choice of modulation and coding, and also depends on the number of hopping channels.  It is assumed that the system uses noncoherent binary continuous-phase frequency-shift keying (CPFSK) modulation, which is the most common choice of modulation for FH systems \cite{cheng:ciss2007}.   The main parameter associated with binary CPFSK is the {\em modulation index}, which characterizes the relative separation between the two tones.  It is furthermore assumed that the system uses a capacity-approaching code (e.g., turbo or LDPC), which allows the achievable performance to be characterized by the capacity of the system, under constraints of the modulation and noncoherent detection technique.  Under the assumption of coded noncoherent CPFSK, the performance of the network is a function of three parameters: the code rate, the modulation index, and the number of hopping channels.

By using the modulation-constrained transmission capacity as the objective function, the paper optimizes the network with respect to these three parameters.  Initially, a brute-force exhaustive optimization is proposed that optimizes over a wide range of discretized parameters.  Because the results of the exhaustive optimization suggest that the optimization problem is convex, a gradient search algorithm \cite{boyd:2004} is proposed that offers a good tradeoff between accuracy and efficiency.

The main contributions of this paper are (1) the closed-form expressions for conditional outage probability and spatially averaged outage probability in the presence of finite-area networks with mobiles drawn from both a BPP and a PPP, (2) the development of {\em modulation constrained} transmission capacity as a performance metric, and (3) a method for optimizing the parameters associated with the ad hoc network.    The methodology presented in this paper presents a new approach to the analysis and optimization of finite ad hoc networks and presents fresh insight into the tradeoffs among the number of frequency-hopping channels, the modulation index, and the code rate in a frequency-hopping network.

\section{Network Model} \label{Section:SystemModel}
The network comprises $M+2$ mobiles 
that include a reference receiver, a reference transmitter $X_{0}$, and $M$ interfering transmitters $X_{1},...,X_{M}.$ The coordinate system is selected such that
the receiving mobile $X_0$ is at the origin.   The variable $X_{i}$ represents
both the $i^{th}$ mobile and its location, and $||X_{i}||$ is the
distance from $X_i$ to the receiving mobile.    While the interferers can be located in any arbitrary region, we assume they are located in an annular region with inner radius $r_{ex}$ and outer radius $r_{net}$. A nonzero $r_{ex}$ may be used to model the effects of interference-avoidance protocols \cite{hasan:2007}.  In particular, a nonzero $r_{ex}$ models an exclusion zone placed around the receiver, which can be realized by having the receiver send a short clear-to-send (CTS) packet in response to a request-to-send (RTS) packet sent by the transmitter.  Under a carrier-sense multiple-access (CSMA) protocol, mobiles within distance $r_{ex}$ from the receiver that overhear the CTS  will suppress their transmission.

$X_{i}$ transmits a signal whose average received power in the absence of fading is $P_{i}$ at a reference distance $r_{0}$.
At the receiving mobile, $X_i$'s  power is
\begin{eqnarray}
  \rho_i
  & = &
  P_i g_i f( ||X_i|| ) \label{eqn:power}
\end{eqnarray}
where  $g_i$ is the power gain due to fading, and $f( ||X_i|| )$ is a path-loss function.  Each $g_i = a_i^2$, where $a_i$ is Rayleigh and $g_i$ unit-mean exponential, i.e. $g_i \sim \mathcal E(1)$. \ For $r\geq r_{0}$, the path-loss function is
expressed as the attenuation power law:
\begin{eqnarray}
   f \left( r \right)
   & = &
   \left( \frac{r}{r_0} \right)^{-\alpha} \label{eqn:pathloss}
\end{eqnarray}
where $\alpha \geq 2$ is the attenuation power-law exponent and $r_0$ is sufficiently large so that the signals are in the far field.

Channel access is through a synchronous frequency-hopping protocol.  The hopping is slow, with multiple symbols per hop, which is a more suitable strategy for ad hoc networks than fast hopping \cite{torrieri:2011}.  An overall frequency band of $B$ Hz is divided into $L$ frequency channels, each of bandwidth $B/L$ Hz.  The transmitters independently select their transmit frequencies with equal probability.  Let $p_i$ denote the probability that interferer $X_i$ selects the same frequency as the source.  Let
$d_i \leq 1$ be the duty factor of the interferer.  It follows that $p_i=d_i/L$ and that using a duty factor less than unity is equivalent to hopping over more than $L$ frequencies \cite{torrieri:2011}.  Assuming that $d_i=d$ for all interferers, $L'=L/d$ denotes the {\em equivalent} number of frequency channels.  It is assumed that the \{$g_{i}\}$ remain fixed for the duration of a hop, but vary independently from hop to hop (block fading).  While the $\{g_{i}\}$ are independent from user to user, they are not necessarily identically distributed.

The instantaneous SINR at the receiving mobile is
\begin{eqnarray}
   \gamma
   & = &
   \frac{ \rho_0 }{ \displaystyle {\mathcal N} + \sum_{i=1}^{M} I_i \rho_i } \label{Equation:SINR1}
\end{eqnarray}
where $\mathcal N$ is the noise power and $I_i$ is a variable that indicates the presence and type of interference (i.e. co-channel interference or adjacent-channel interference).   When adjacent-channel interference \cite{torrieri:2011} is neglected, $I_i=1$ when $X_i$ selects the same frequency as $X_0$, and $I_i=0$ otherwise.  It follows that $I_i$ is Bernoulli with probability $P[I_i=1]=p_i$.

Substituting (\ref{eqn:power}) and (\ref{eqn:pathloss}) into (\ref{Equation:SINR1}),  the SINR is
\begin{eqnarray}
   \gamma
   & = &
   \frac{ g_0 \Omega_0^{-1}  }{ \displaystyle \Gamma^{-1} + \sum_{i=1}^M I_i g_i \Omega_i^{-1} }
   \label{Equation:SINR2}
\end{eqnarray}
where $\Gamma = r_0^\alpha P_{0}/\mathcal{N}$ is the signal-to-noise
ratio (SNR)  when the transmitter is at unit distance and fading  is absent, $\Omega_i = (P_0/P_i)||X_i||^{\alpha}$ is the inverse normalized power of $X_i$ at the receiver, and $\Omega_0 = ||X_0||^{\alpha}$.  Without loss of generality\footnote{Changing $||X_0||$ is equivalent to a scaling of $r_{ex}$ and $r_{net}$.}, we assume that $||X_0|| = 1$ for the remainder of this paper.  Furthermore, all examples and numerical results in this paper assume that all mobiles transmit with the same power, i.e. $P_i = P_0$ for all $i$.

\section{Conditional Outage Probability} \label{Section:Outage}
Let $\beta$ denote the minimum SINR required for reliable reception and $\boldsymbol{\Omega }=\{\Omega_{0},...,\Omega _{M}\}$ represent the set of inverse normalized powers.  An \emph{outage} occurs when the SINR falls below $\beta$.  Conditioning on $\boldsymbol{\Omega }$, the outage probability is
\begin{eqnarray}
   \epsilon_{\Omega}
   & = &
   P \left[ \gamma \leq \beta \big| \boldsymbol \Omega \right].
   \label{Equation:Outage1}
\end{eqnarray}
Because it is conditioned on $\boldsymbol{\Omega }$, the outage probability depends on the particular network geometry, which has dynamics over timescales that are much slower than the fading. By defining a variable
\vspace{-0.5cm}
\begin{eqnarray}
 \mathsf Z_{M} & = & \beta^{-1} g_0 \Omega_0^{-1} - \sum_{i=1}^M g_i I_i \Omega_i^{-1} \label{eqn:z}
\end{eqnarray}
the conditional outage probability may be expressed as
\begin{eqnarray}
  \epsilon_{\Omega}
  & = &
  P
  \left[
   \mathsf Z_{M}  \leq \Gamma^{-1} \big| \boldsymbol \Omega
  \right]
  = F_{\mathsf Z_{M}} \left( \Gamma^{-1} \big| \boldsymbol \Omega \right) \label{Equation:OutageCDF}
\end{eqnarray}
which is the cumulative distribution function (cdf) of $\mathsf Z_M$ conditioned on $\boldsymbol \Omega$ and evaluated at $\Gamma^{-1}$.

By defining $\mathsf S = \beta^{-1} \Omega_0^{-1} g_0$ and $\mathsf Y_i =I_i g_i \Omega_i^{-1}$, (\ref{eqn:z}) may be rewritten as
\vspace{-0.5cm}
\begin{eqnarray}
   \mathsf Z_{M}
   & = &
   \mathsf S - \sum_{i=1}^M \mathsf Y_i.
\end{eqnarray}
where $\mathsf S \sim {\mathcal E}( \beta \Omega_0 )$.  The cdf of $\mathsf S$ is
\begin{eqnarray}
   F_{\mathsf S}(y)
   & = &
   \left( 1 - e^{-\beta \Omega_0 y} \right) u(y) \label{cdf}
\end{eqnarray}
where $u(y)$ is the unit-step function.
Taking into account the Rayleigh fading and Bernoulli $\{I_i\}$, the pdf of ${\mathsf Y_i}$ is
\begin{eqnarray}
  f_{\mathsf Y_i}(y)
   & = &
  (1-p) \delta(y)
  + p_i \Omega_i
   e^{- \Omega_i y }
  u(y)
  \label{pdf}
\end{eqnarray}
where $\delta(y)$ is the Dirac delta function.

First consider the single-interferer case ($M=1$).  The cdf of $\mathsf Z_1$ is expressed as
\vspace{-0.25cm}
\begin{eqnarray}
F_{\mathsf Z_1}(z \big| \boldsymbol \Omega)
& = &
 \int_{0}^{\infty}F_{\mathsf S}(z+y)f_{\mathsf Y_1}(y) dy. \label{op3}
\end{eqnarray}
Substituting (\ref{cdf}) and (\ref{pdf}) into (\ref{op3}) yields
\begin{eqnarray}
F_{\mathsf Z_1}(z \big| \boldsymbol \Omega)
& = &
\int_{0 }^{\infty} \left[ 1 - e^{-\beta \Omega_0(z + y) } \right]
f_{\mathsf Y_1}(y)  dy \nonumber \\
& = &
1 - e^{-\beta \Omega_0 z } \left( \frac{(1-p_1)\beta \Omega_0 + \Omega_1 }{\beta \Omega_0 +\Omega_1} \right) \label{op4}
\end{eqnarray}
for $z \geq 0$. Using the fact that $\mathsf Z_{M} = \mathsf Z_{M-1} - \mathsf Y_M$, and working iteratively
\vspace{-0.25cm}
\begin{multline}
F_{\mathsf Z_M}(z \big|\boldsymbol \Omega)  =  1 -
 e^{-\beta \Omega_0 z }
\prod_{i=1}^M
\left[ \frac{(1-p_i)\beta \Omega_0 +\Omega_i}{\beta \Omega_0+\Omega_i} \right]
 \label{op7}
\end{multline}
for $z \geq 0$.  The outage probability is found by substituting $z=\Gamma^{-1}$ into the above expression.

\begin{figure}[t]
\centering
\hspace{-0.5cm}
\includegraphics[width=9.25cm]{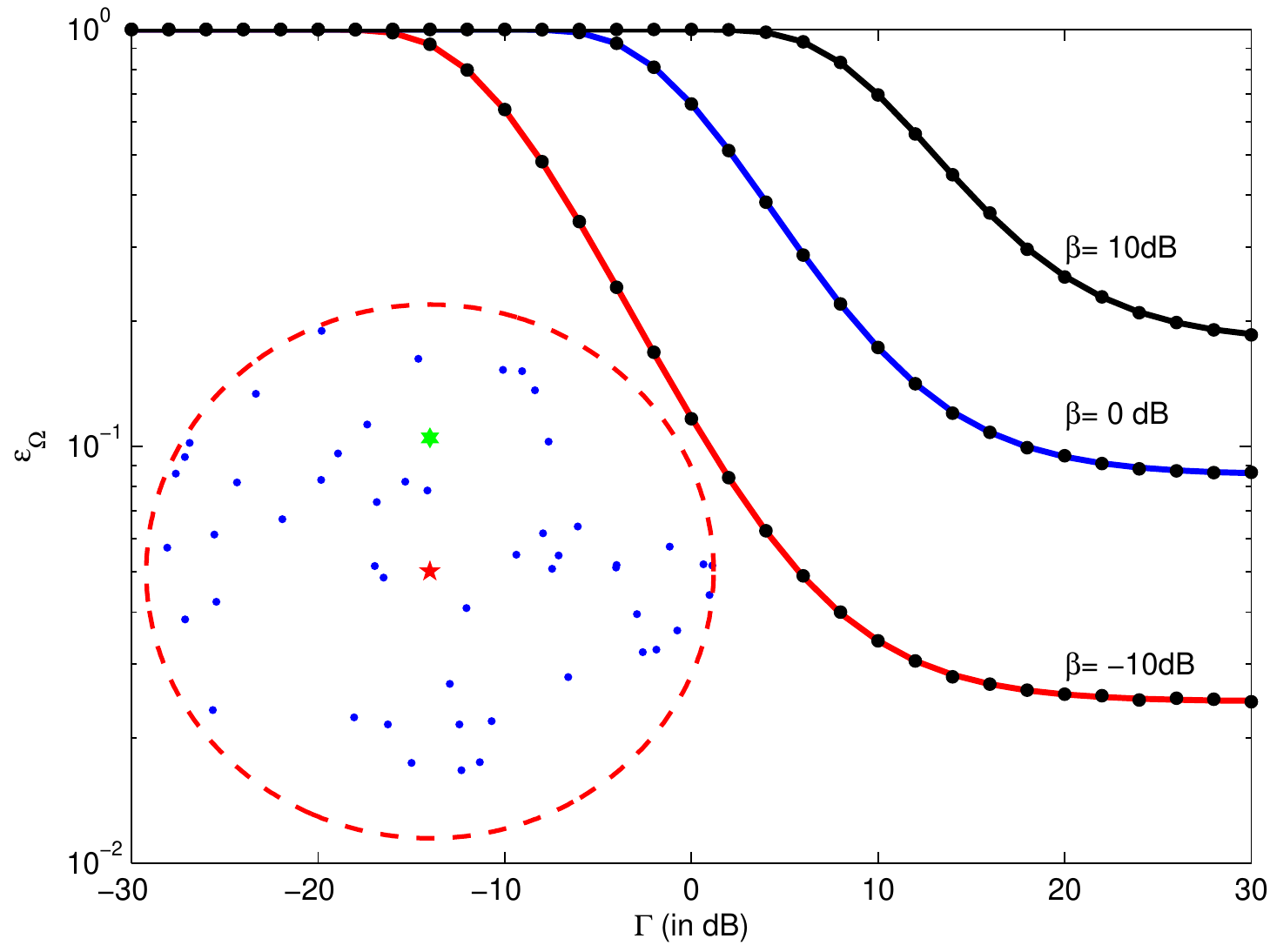}
\vspace{-0.5cm}
\caption{Conditional outage probability $\epsilon_{\Omega}$ as a function of SNR $\Gamma$.  Analytical curves are solid, while dots represent simulated values.  Top curve: $\beta=10$ dB. Middle curve: $\beta=0$ dB.  Bottom curve: $\beta=-10$ dB. The network geometry is shown in the inset. The receiving mobile is represented by the five-pointed star at the center of the network, the desired transmitting mobile by the six-pointed star immediately above the transmitter, and the 50 interferers are shown as dots.  \label{Figure:Example1} }
\vspace{-0.5cm}
\end{figure}

{\bf Example \#1:}
Consider a specific network topology with the transmitting mobile placed one unit North of the receiving mobile, and fifty interferers arbitrarily placed in an annular region with outer radius $r_{net}=2$ and inner radius $r_{ex}=0.25$. The resulting network is shown in the inset of Fig. \ref{Figure:Example1}.  The $\boldsymbol \Omega$ was determined by assuming a path-loss exponent $\alpha = 3$ and a common transmit power $P_i=P_0$. The equivalent number of frequency channels was set to $L'=200$.
Fig. \ref{Figure:Example1} shows the outage probability as a function of the SNR $\Gamma$, computed at each SNR point by evaluating (\ref{op7}) at $z=\Gamma^{-1}$. Three cases were considered for the SINR threshold: $\beta= -10$ dB, $\beta= 0$ dB and $\beta= 10$ dB. Also shown are results generated by simulation, which involved randomly generating the exponentially-distributed  $\{g_i\}$. The analytical and simulation results coincide, which is what is to be expected because  (\ref{op7}) is exact.  Any discrepancy between the curves can be attributed to the finite number of Monte Carlo trials (one million trials were executed per SNR point).

\section{Outage of a BPP}
\label{Section:BPP}
Because it is conditioned on ${\boldsymbol \Omega}$, the outage probability $\epsilon_\Omega$ presented in the last section depends on the geometry of the particular network, i.e., the location of the interferers.  The conditioning on ${\boldsymbol \Omega}$ can be removed by averaging ${F}_{\mathsf Z_M}(z|\boldsymbol \Omega)$ over the spatial distribution of the network.  In a BPP, a fixed number $M$ of mobiles are independently and uniformly distributed over the network.  Let $\epsilon_M$ be the spatially averaged outage probability when the interferers are drawn from a BPP, which is found by taking the expectation of ${F}_{\mathsf Z_M}(z|\boldsymbol \Omega)$ with respect to ${\boldsymbol \Omega}$:
\begin{eqnarray}
   \epsilon_M
   &=&
   E_{\boldsymbol \Omega} \left[
   \epsilon_\Omega
   \right]
   =
   E \left[
   {F}_{\mathsf Z_M}\left( \Gamma^{-1} \big| \boldsymbol \Omega \right)
   \right]
   =
   {F}_{\mathsf Z_M}\left( \Gamma^{-1} \right). \nonumber \\
\end{eqnarray}
In the above equation, ${F}_{\mathsf Z_M}\left( z \right)$ is the cdf of $\mathsf Z_M$ averaged over the spatial distribution, which can be found analytically or computed through Monte Carlo simulation.  To compute it via Monte Carlo simulation, generate a large number $N$ of networks, each containing $M$ interferers drawn from a BPP.  Compute the ${F}_{\mathsf Z_M}(z|\boldsymbol \Omega)$ for each network by using the method outlined in Section III, and average over the $N$ networks.  Letting $\boldsymbol \Omega_n$ be normalized inverse power coefficients of the $n^{th}$ randomly generated network, the Monte Carlo estimate of the cdf is
\begin{eqnarray}
   {F}_{\mathsf Z_M}(z)
   & = &
   \frac{1}{N}
   \sum_{n=1}^N
   {F}_{\mathsf Z_M}(z|\boldsymbol \Omega_n).\label{Equation:MC}
\end{eqnarray}
Note that the Monte Carlo simulation only requires the realization of the interferer locations, and does not require the realization of the fading coefficients.

For a BPP constrained to an annular region with inner radius $r_{ex}$ and outer radius $r_{net}$, the spatial coordinates can be represented as the complex value $X_i=r_i e^{j \theta_i}$.  The location can then be realized by drawing two independent numbers $x_{1,i}$ and $x_{2,i}$ from the uniform distribution over $\left[\left(\frac{r_{ex}}{r_{net}}\right)^2,1 \right]$ and $[0,1]$ respectively and then setting $r_i= r_{net} \sqrt{x_{1,i}}$ and $\theta_i=2 \pi x_{2,i}$.

To avoid the computational burden of a Monte Carlo simulation, a closed-form expression for the spatially averaged outage probability is preferred.  Since $||X_i||=r_i$ and $\Omega_i=||X_i||^\alpha$, it follows that the normalized inverse power of the $i^{th}$ interferer is
\begin{eqnarray}
\Omega_i &=& \left( \sqrt{x_{1,i}} r_{net} \right)^{\alpha}.
\end{eqnarray}
The pdf of $\Omega_i$ is
\begin{eqnarray}
f_{\Omega_i}(\omega) &=&  \frac{2}{\alpha} \omega^{\frac{2-\alpha}{\alpha}} \left( r_{net}^2- r_{ex}^2 \right)^{-1}
\label{pdf_omega}
\end{eqnarray}
for $r_{ex}^{\alpha} \leq \omega \leq r_{net}^\alpha$, and zero otherwise.  The spatially averaged outage probability can now be obtained using (\ref{op7}) and ($\ref{pdf_omega}$) as follows:
\begin{eqnarray}
F_{\mathsf Z_M}(z) &=&  \int  f_{\boldsymbol \Omega}(\boldsymbol \omega) F_{\mathsf Z_M}(z \big| \boldsymbol \omega) d\boldsymbol \omega
\label{cdf_M}
\end{eqnarray}
where the $M$-fold integral is over the joint pdf of $\{\Omega_1, ..., \Omega_M\}$.
Substituting (\ref{op7}) and ($\ref{pdf_omega}$) into ($\ref{cdf_M}$) and using the fact that the $\{\Omega_i\}$ are independent yields:
\begin{multline}
F_{\mathsf Z_M}(z)  =  1 - e^{-\beta \Omega_0 z }  \prod_{i=1}^M \frac{2}{\alpha} \left( r_{net}^2- r_{ex}^2 \right)^{-1} \\
\int_{r_{ex}^{\alpha} }^{r_{net}^{\alpha} }
\omega^{\frac{2-\alpha}{\alpha}} \left( \frac{(1-p_i)\beta \Omega_0 +\omega}{\beta \Omega_0+\omega} \right) d\omega
\label{cdf_M_1}
\end{multline}
Evaluating the integral and assuming $p_i=p$ for all users results in:
\begin{eqnarray}
F_{\mathsf Z_M}(z)= 1 - e^{-\beta_0  z } \kappa^M \left \{ \Psi \left( r_{net}^{\alpha} \right) - \Psi \left( r_{ex}^{\alpha} \right) \right \}^M
\label{cdf_BPP}
\end{eqnarray}
where $\beta_0= \beta \Omega_0$,
\begin{eqnarray}
\kappa & = & \left( r_{net}^2- r_{ex}^2 \right)^{-1} \label{kappa}
\end{eqnarray}
\begin{multline}
\Psi(x)  =  x^{\frac{2}{\alpha}} \cdot (1-p) + \frac{2 \cdot p }{\alpha+2} \cdot \frac{x^{\frac{2+\alpha}{\alpha}}}{\beta_0} \\
   \times {_2F}_1\left( \left[1, \frac{\alpha+2}{\alpha}\right]; \frac{2\alpha+2}{\alpha}, -\frac{x}{\beta_0}\right)   \label{Psi}
\end{multline}
and $_2F_1([a,b];c,x)$  is the Gauss hypergeometric function given by \cite{Abramowitz:1965}
\vspace{-0.5cm}
\begin{multline}
 _2F_1 ([a,b];c;x) = \frac{\Gamma(c)}{\Gamma(b)\Gamma(c-b)} \\
 \times
  \int_{0 }^{1} \nu^{b-1}(1-\nu)^{c-b-1}(1-\nu x)^{-a}d \nu
\end{multline}
\vspace{-0.25cm}
where
\vspace{-0.25cm}
\begin{eqnarray}
  \Gamma(z) & = & \int_{0 }^{\infty} t^{z-1} e^{-t} dt.
\end{eqnarray}

While the results in this section are for an annular network centered upon the reference receiver, we note that other network shapes can be accommodated by determining the appropriate pdf of the $\Omega_i$ and substituting into (\ref{cdf_M}).  Furthermore, shadowing can be accommodated by using appropriately defined  $f_{\Omega_i}(\omega)$.

\begin{figure}[t]
\centering
\hspace{-0.5cm}
\includegraphics[width=9.25cm]{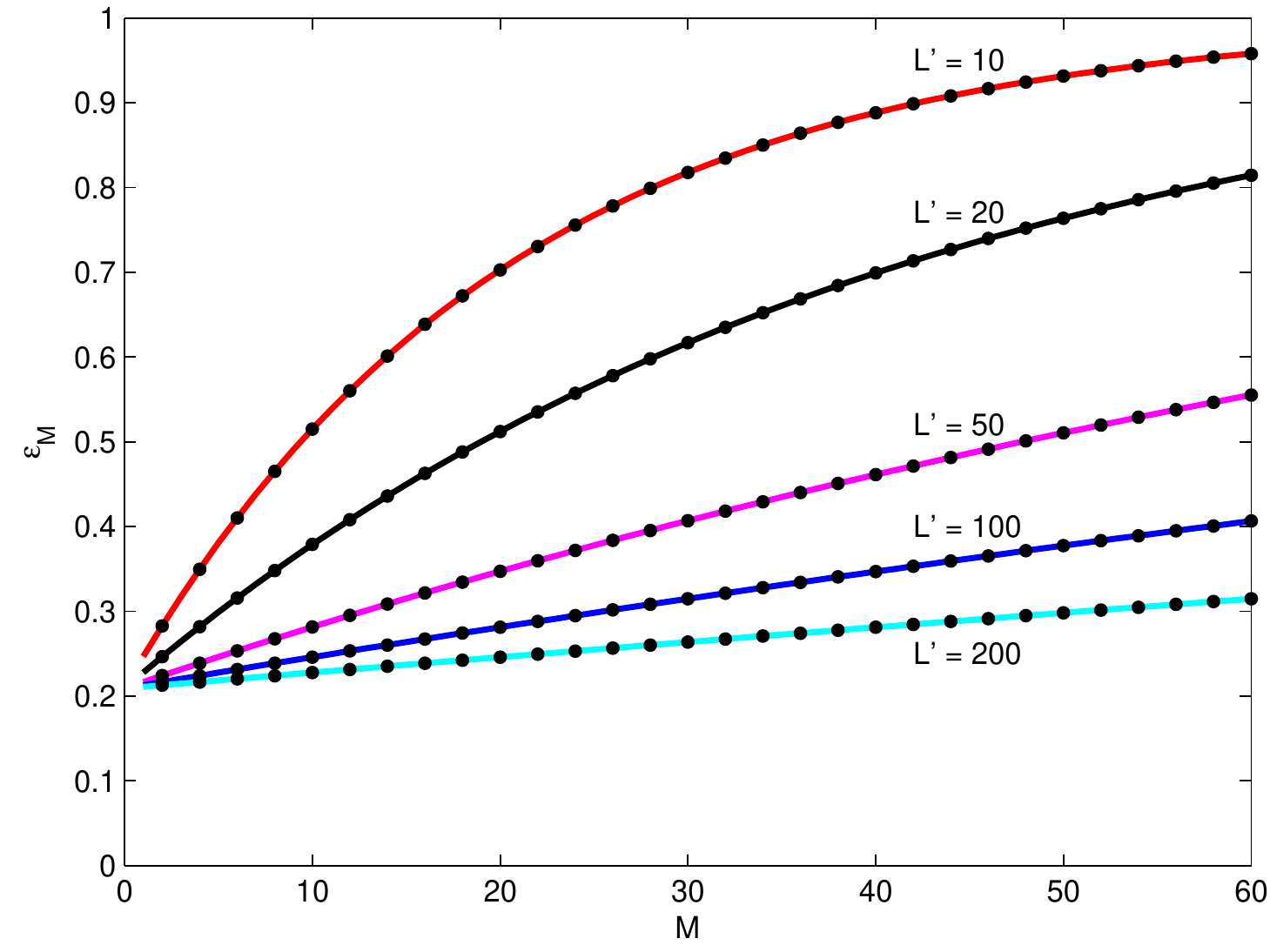}
\vspace{-0.5cm}
\caption{Outage probability $\epsilon_M$ as a function of $M$ for five values $L'$ for networks with $M$ interfering mobiles drawn from a BPP. The  SINR threshold is $\beta = 3.7$ dB, the SNR is set to $\Gamma = 10$ dB, and the other parameters are identical to those used to generate Fig. \ref{Figure:Example1}.  Analytical curves are solid, while dots represent simulated values.    \label{Figure:Example2} }
\vspace{-0.5cm}
\end{figure}

{\bf Example \#2:}
Reconsider Example \#1, but now instead of the network assuming the specific topology shown in Fig. \ref{Figure:Example1}, let the $M$ interferers be placed according to a BPP.  By using (\ref{cdf_BPP}), the spatially averaged outage probability can be found.  Fig. \ref{Figure:Example2} shows the outage probability as a function of $M$ for five values of $L'$ when the SINR threshold is set to $\beta = 3.7$ dB and the SNR  is set to $\Gamma = 10$ dB.  The values of $r_{ex}$, $r_{net}$, and $\alpha$ are the same as in Example \#1.  The solid curves show the spatially averaged outage probability evaluated analytically, i.e., by using (\ref{cdf_BPP}), while the dots show the probability found by Monte Carlo averaging, i.e, by using (\ref{Equation:MC}) with $N=10 \, 000$ randomly generated networks.  Because  (\ref{cdf_BPP}) is exact and $N$ large, the analytical and simulation results coincide. From Fig. \ref{Figure:Example2} it is observed that the outage probability degrades with increasing $M$ and decreasing $L'$.

\begin{figure}[t]
\centering
\hspace{-0.5cm}
\includegraphics[width=9.25cm]{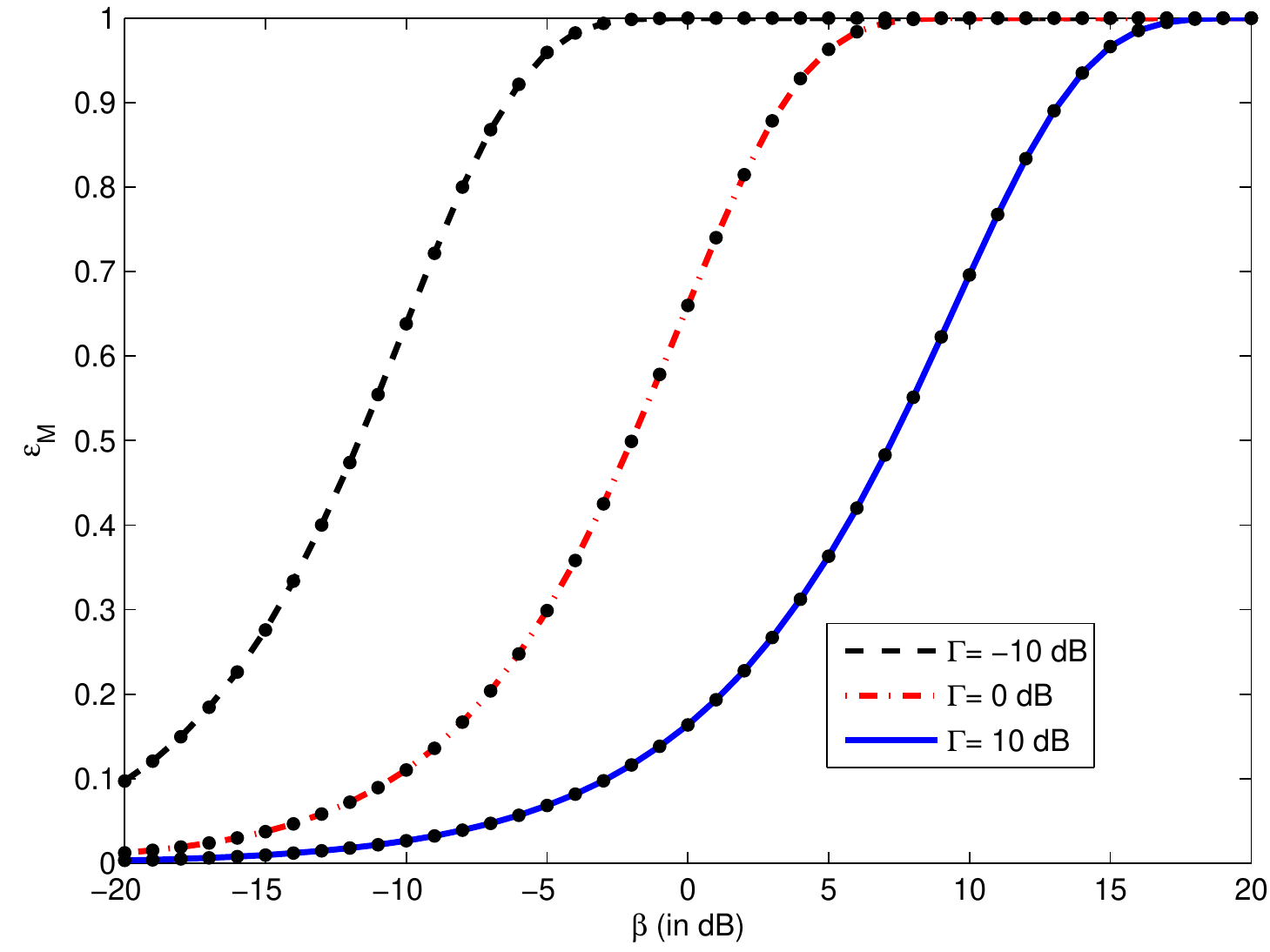}
\vspace{-0.5cm}
\caption{Outage probability $\epsilon_M$ as a function of $\beta$ for three values of $\Gamma$ for an annular network area with outer radius of $r_{net}=2$, inner radius of $r_{min}=0.25$. A fixed number of interfering mobiles ($M = 50$) is drawn from a BPP and a path-loss exponent $\alpha=3$ is fixed. Analytical curves are solid, while dots represent simulated values.    \label{Figure:Example3} }
\vspace{-0.5cm}
\end{figure}

{\bf Example \#3:}  The dependence on $\beta$ and $\Gamma$ is investigated in Fig. \ref{Figure:Example3}.  The values of $r_{ex}$, $r_{net}$, and $\alpha$ are the same as in Example \#1.  The number of interferers is set to $M=50$ and the equivalent number of frequency channels set to $L' = 200$.  As in Example \#2, the spatially averaged outage probability is computed analytically and through simulation.  In particular, the solid curves show the spatially averaged outage probability evaluated using (\ref{cdf_BPP}), while the dots show the probability found by Monte Carlo averaging with $N=10 \, 000$ randomly generated networks.  The outage probability is shown as a function of the SINR threshold $\beta$ for three values of $\Gamma$.  Again the analytical curves coincide with the simulation results.  From  \ref{Figure:Example3}, it is observed that the outage probability increases with increasing $\beta$ and decreases with increasing $\Gamma$.

\section{Outage of a PPP}
\label{Section:Poisson Point Process in a Finite Network}
Suppose that the network now has a variable number of interferers $M$.  Let $p_M(m)$ indicate the probability mass function (pmf) of $M$.  Let $F_{\mathsf Z_m}(z)$ be the cdf of $Z_m$ when there are $m$ interferers drawn from a BPP.  It follows that the spatially averaged cdf for the variable-sized network is
\begin{eqnarray}
F_{\mathsf Z}(z)
& = &
 \sum_{ m=0}^{ \infty } p_{M}(m) F_{\mathsf Z_m}(z)
\label{cdf_PPP}
\end{eqnarray}
and the outage probability averaged over a spatial distribution with a variable number of interferers is $\epsilon  = E[ \epsilon_M ] = F_{\mathsf Z}\left( \Gamma^{-1} \right)$, where the expectation is with respect to the distribution of $M$ as given by (\ref{cdf_PPP}).
We note that the $\{ F_{\mathsf Z_M}(z) \}$ in (\ref{cdf_PPP}) may be obtained either through simulation using (\ref{Equation:MC}), which requires the summation in (\ref{cdf_PPP}) to be truncated.  Alternatively, the analytical expression given by (\ref{cdf_BPP}) may be used, which as will be shown below, does not require truncation of the summation.

When the spatial distribution is a PPP, the distribution of $M$ is Poisson with intensity $\eta=\lambda A$, where $\lambda$ is the density of the points per unit area and $A$ is the area over which the points are distributed.
For a PPP of density $\lambda$, the number of interfering mobiles $M$ within area $A$ has pmf
\vspace{-0.2cm}
\begin{eqnarray}
p_{M}(m)
& = & \frac{(\lambda A)^{m}}{m!}e^{-\lambda A}
\label{pmf_Poisson}
\end{eqnarray}
for $m\geq 0$.

\noindent Substituting ($\ref{cdf_BPP}$) and ($\ref{pmf_Poisson}$) into ($\ref{cdf_PPP}$) yields:
\vspace{-0.25cm}
\begin{multline}
F_{\mathsf Z}(z)= e^{-\lambda A}\sum_{ m=0}^{ \infty } \frac{(\lambda A)^{m}}{m!} \\
 \left\{ 1 - e^{-\beta_0  z }  \left \{ \kappa \cdot \left[ \Psi \left( r_{net}^{\alpha} \right) - \Psi \left( r_{ex}^{\alpha} \right)\right] \right \}^m \right\}
\label{cdf_PPP_2}
\end{multline}
where $\kappa$ and $\Psi(x)$ are given  by ($\ref{kappa}$) and ($\ref{Psi}$), respectively.
By using the identity \cite{Gradshteyn:2007}
\vspace{-0.2cm}
\begin{eqnarray}
\sum_{ m=0}^{ \infty } \frac{a^m}{m!}\left( 1-c \cdot b^m \right)= e^a-c \cdot e^{a \cdot b}
\end{eqnarray}
and $A = \pi \left( r_{net}^2 - r_{ex}^2 \right )$, (\ref{cdf_PPP_2}) may be expressed as
\begin{eqnarray}
F_{\mathsf Z}(z) & = & 1 - \exp \left\{ -\beta_0  z   - \pi \lambda \left( r_{net}^2 - r_{ex}^2 \right)  \right. \nonumber \\
& & \left. \times \left ( 1- \kappa \left[ \Psi \left( r_{net}^{\alpha} \right) - \Psi \left( r_{ex}^{\alpha} \right)\right] \right ) \right\}.
\label{PPPNN}
\end{eqnarray}


The expression given in (\ref{PPPNN}) generalizes an earlier expression given in Baccelli et al. \cite{baccelli:2006} for the outage probability of an infinite network ($r_{net} \rightarrow \infty$) with no exclusion zone ($r_{ex}=0$) and constantly transmitting mobiles ($p=1$).  To see this, set $p=1$ and $r_{ex}=0$  in (\ref{PPPNN}) and take the limit as $r_{net} \rightarrow  \infty$,
\vspace{-0.3cm}
\begin{multline}
F_{\mathsf Z}(z)  =  \lim_{r_{net} \rightarrow   \infty  }  1 - \exp \left\{-\beta_0  z - \pi \lambda r_{net}^2
  \left[ 1- \frac{r_{net}^{\alpha}}{\beta_0} \right. \right. \\
 \left.  \left. \frac{2 }{(\alpha+2)} \cdot {_2F}_1\left( \left[1, \frac{\alpha+2}{\alpha}\right]; \frac{2\alpha+2}{\alpha}, -\frac{r_{net}^{\alpha}}{\beta_0}\right) \right] \right\}.
\label{Limit2}
\end{multline}
By using the identity \cite{Abramowitz:1965}
\vspace{-0.2cm}
\begin{eqnarray}
 {_2F}_1\left( \left[a, b \right]; c, z \right) = (1-z)^{-b}  {_2F}_1\left( \left[ b, c-a \right]; c, \frac{z}{z-1} \right)
\label{identity}
\end{eqnarray}
 and performing some algebraic manipulations, (\ref{Limit2}) becomes
\vspace{-0.4cm}
\begin{multline}
  F_{\mathsf Z}(z)=  1 - \lim_{r_{net} \rightarrow   \infty  } \exp \left \{-\beta_0  z  -\frac{2 \pi \lambda r_{net}^{\alpha+2}}{\beta_0 (\alpha+2)}
  \left( \frac{r_{net}^{\alpha}}{\beta_0}+1 \right)^{- \frac{\alpha+2}{\alpha}} \right. \\
  \left. {_2F}_1\left( \left[ \frac{\alpha+2}{\alpha}, \frac{\alpha+2}{\alpha}\right]; \frac{2\alpha+2}{\alpha}, \frac{\frac{r_{net}^{\alpha}}{\beta_0}}{\frac{r_{net}^{\alpha}}{\beta_0}+1} \right) \right \}.
\label{Limit4}
\end{multline}
Because $\frac{r_{net}^{\alpha}}{\beta_0}+1 =\frac{r_{net}^{\alpha}}{\beta_0}$ when $r_{net} \rightarrow  \infty$, (\ref{Limit4}) can be simplified to
\vspace{-0.4cm}
\begin{multline}
  F_{\mathsf Z}(z)=  1 - \lim_{r_{net} \rightarrow   \infty  } \exp \left \{-\beta_0  z -\frac{2 \pi \lambda}{(\alpha+2)} \beta_0^{\frac{2}{\alpha}} \right. \\
  \left. {_2F}_1\left( \left[ \frac{\alpha+2}{\alpha}, \frac{\alpha+2}{\alpha}\right]; \frac{2\alpha+2}{\alpha}, \frac{\frac{r_{net}^{\alpha}}{\beta_0}}{\frac{r_{net}^{\alpha}}{\beta_0}+1} \right) \right\}.
\label{Limit5}
\end{multline}
Since $ \displaystyle \lim_{r_{net} \rightarrow   \infty  } \frac{ r_{net}^{\alpha}/ \beta_0 }{ r_{net}^{\alpha} / \beta_0 + 1} = 1$, (\ref{Limit5}) is equal to
\vspace{-0.2cm}
\begin{multline}
  F_{\mathsf Z}(z)=  1 -\exp \left \{-\beta_0  z -\frac{2 \pi \lambda}{(\alpha+2)} \beta_0^{\frac{2}{\alpha}} \right. \\
  \left. {_2F}_1\left( \left[ \frac{\alpha+2}{\alpha}, \frac{\alpha+2}{\alpha}\right]; \frac{2\alpha+2}{\alpha}, 1 \right) \right\}.
\label{Limit6}
\end{multline}
By using the identity \cite{Abramowitz:1965}
\begin{eqnarray}
 {_2F}_1\left( \left[a, b \right]; c, 1 \right)&  = & \frac{\Gamma(c)\Gamma(c-a-b)}{\Gamma(c-a)\Gamma(c-b)}
\label{identity1}
\end{eqnarray}
and performing a few algebraic manipulations, (\ref{Limit6}) becomes
\begin{multline}
F_{\mathsf Z}(z)
 =
1 -\exp \left \{-\beta_0  z -\frac{2 \pi \lambda}{{\alpha}}\beta_0^{\frac{2}{\alpha}} \Gamma\left(\frac{2}{\alpha}\right) \Gamma\left(1-\frac{2}{\alpha}\right)\right \}.
\label{Baccelli}
\end{multline}
which, in the absence of noise, coincides with equation (3.4) of \cite{baccelli:2006} and equation (61) of \cite{weber:2010}.

\begin{figure}[t!]
\centering
\hspace{-0.5cm}
\includegraphics[width=9.25cm]{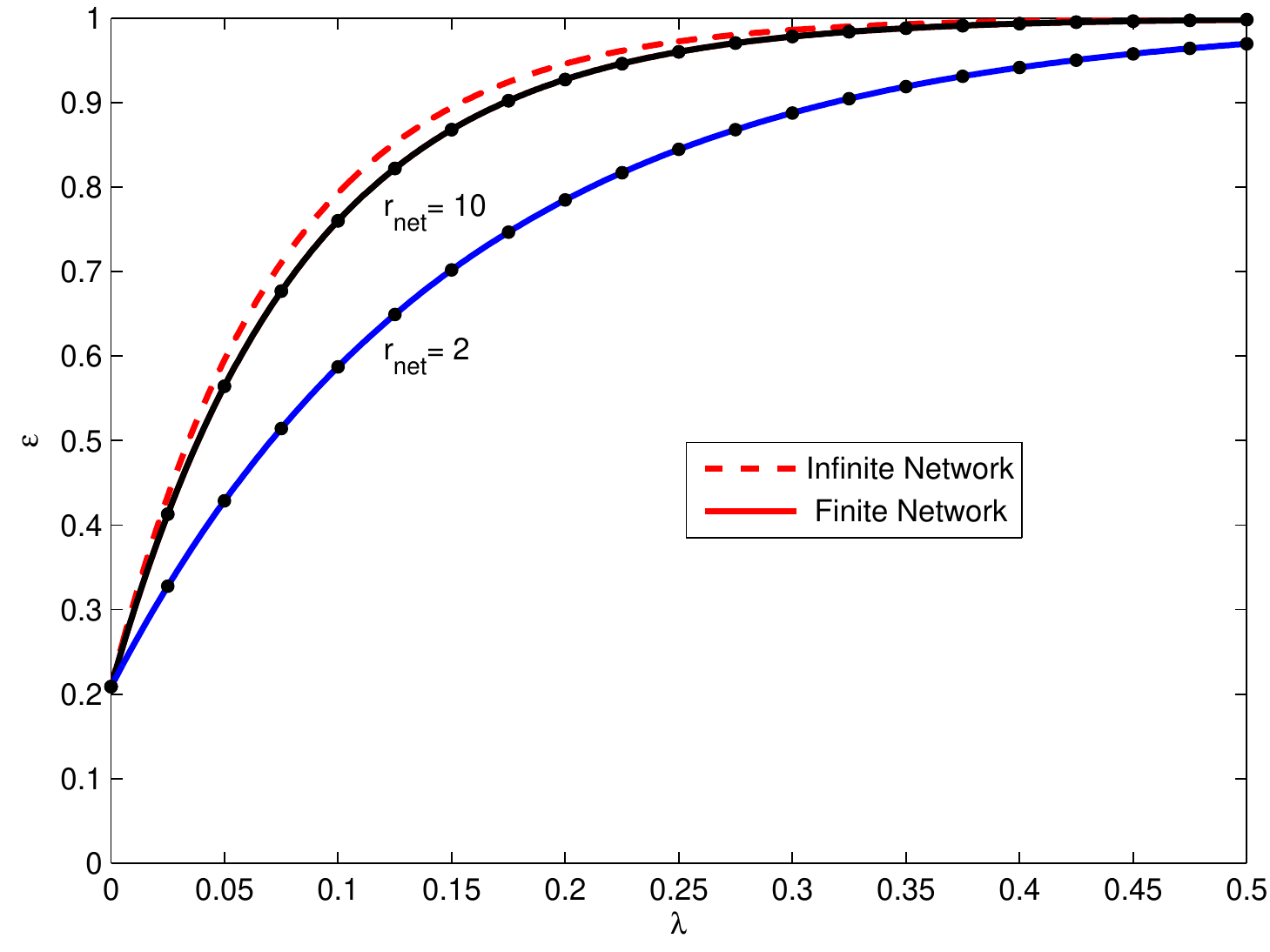}
\vspace{-0.5cm}
\caption{Spatially averaged outage probability $\epsilon$ as a function of the mobile density $\lambda$ when interferers are drawn from a PPP. Analytical curves are solid, while dots represent simulated values.  Top solid curve: $r_{net}=10$.  Bottom solid curve: $r_{net}=2$. The dotted line curve is the $\epsilon$ as a function of $\lambda$ for an infinite network. \label{Figure:Example4} }
\vspace{-0.5cm}
\end{figure}

{\bf Example \#4:}  As with Example \#2, suppose that $||X_{0}|| = 1$, $\alpha = 3$, and $\Gamma=10$ dB. Let $r_{ex}=0$ and $L'=1$.  The interfering mobiles are now placed in a circular region of radius $r_{net}$ according to a PPP with node density $\lambda$.  Three network radii are considered: $r_{net} = \{2, 10, \infty\}$. The SINR threshold is set to $\beta=3.7$ dB.  Fig. \ref{Figure:Example4} shows the spatially averaged outage probabilities for each of the three values of $r_{net}$ as a function of $\lambda$.  The outage probabilities of the two networks with finite radius were computed using (\ref{PPPNN}), while the outage probability of the infinite network was computed using ($\ref{Baccelli}$).  In addition, simulation results are shown for the two finite networks, which coincide with the theoretical curves.  Again, one million trials were executed per $\lambda$ point, and any discrepancy between the theoretical result and the simulation is due to the finite number of trials.
Fig. \ref{Figure:Example4} shows that the outage probability increases with increasing $\lambda$ and/or increasing $r_{net}$.

\begin{figure}[t!]
\centering
\hspace{-0.5cm}
\includegraphics[width=9.25cm]{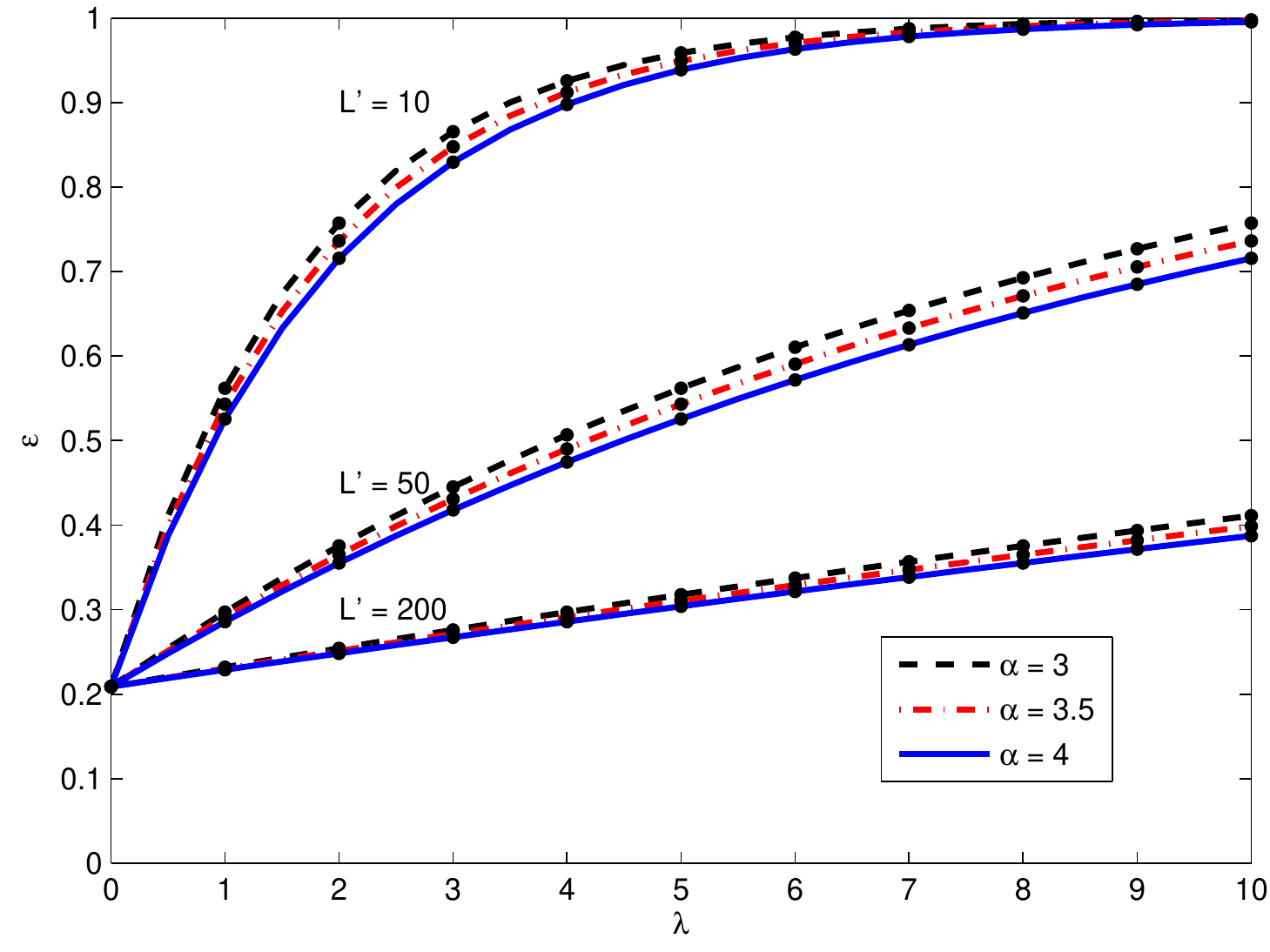}
\vspace{-0.5cm}
\caption{Spatially averaged outage probability $\epsilon$ as a function of node density $\lambda$ for three values $L'$ when interferers are drawn from a PPP. The source transmitter is placed at unit distance from the reference receiver and a exclusion-zone is imposed at the receiver with radius $r_{ex}=0.25$. The SINR threshold is fixed to $\beta = 3.7$ dB and the SNR is set to $\Gamma = 10$.
 Analytical curves are solid, while dots represent simulated values. \label{Figure:Example5} }
\vspace{-0.5cm}
\end{figure}

{\bf Example \#5:}  Fig.  \ref{Figure:Example5}  investigates the influence of the path-loss exponent $\alpha$ and the number of equivalent hopping channels $L'$.  As in Example \#4, $||X_{0}|| = 1$ and $\Gamma=10$ dB. The interferering mobiles are placed in an annular region with an inner radius $r_{ex}=0.25$ and an outer radius $r_{net}=2$ according to a PPP with node density $\lambda$, and the SINR threshold is fixed to $\beta=3.7$ dB. Fig. \ref{Figure:Example5} shows the outage probability as a function of $\lambda$ for three values of $L'$ and three values of path-loss exponent $\alpha$. For each set of ($L',\alpha$), the outage probability was computed analytically using (\ref{PPPNN}), as shown by the curves, and by Monte Carlo simulation with one million trials, as indicated by the dots.  Consistent with observations made in Examples \#2 and \#4, Fig. \ref{Figure:Example5} shows that the outage probability increases with decreasing $L'$ and increases with increasing node density.  In addition, Fig. \ref{Figure:Example5} shows that the outage probability decreases with increasing path-loss exponent $\alpha$.

\section{Transmission Capacity}
Often, networks are constrained to ensure that the outage probability $\epsilon$ does not exceed a maximum outage probability $\zeta$ $\in$ $\left[ 0,1 \right]$; i.e., $\epsilon \leq \zeta$.  Under such a constraint, the maximum density of transmissions is of interest, which is quantified by the {\em transmission capacity} (TC) \cite{weber:2010}.  With outage constraint $\zeta$, the TC is
\begin{eqnarray}
\tau_c\left(\zeta \right)
 & = &
\epsilon^{-1}(\zeta)(1-\zeta)
\label{TC_definition}
\end{eqnarray}
where $\epsilon^{-1}(\zeta)$ is the density of the underlying process (BPP or PPP) whose spatially averaged outage probability satisfies the constraint $\epsilon \leq \zeta$ with equality\footnote{Since $\epsilon$ is a monotonically increasing function of $\lambda$, the TC is maximized when the constraint $\epsilon \leq \zeta$ is met with equality.}, and $(1-\zeta)$ ensures that only successful transmissions are counted.   The TC represents the spatial spectral efficiency; i.e. the rate of successful data transmission per unit area.  With appropriately normalized variables, the TC can assume units of bits-per-second per Hz per $m^2$ (bps/Hz/$m^2$).

Closed-form expressions for TC can be found in Rayleigh fading for networks drawn from either a BPP or a PPP.  For the BPP case, $\epsilon^{-1}(\zeta)$ is found be solving $\epsilon =  F_{\mathsf Z_M}( \Gamma^{-1} ) = \zeta$ for $\lambda$.  By substituting $M = \lambda A$ into (\ref{cdf_BPP}),
\begin{eqnarray}
\zeta
& = &
 1 - e^{-\beta_0 \Gamma^{-1}} \left[ \kappa \left \{ \Psi \left( r_{net}^{\alpha} \right) - \Psi \left( r_{ex}^{\alpha} \right) \right \} \right]^{\lambda A}\hspace{-0.4cm}.
\end{eqnarray}
By solving for $\lambda$ and setting the result to $\epsilon^{-1}(\zeta)$,
\begin{eqnarray}
\epsilon^{-1}(\zeta)
& = &
\frac{ \log( 1 - \zeta ) + \beta_0 \Gamma^{-1} }{ A   \log \left\{ \kappa \left[ \Psi \left( r_{net}^{\alpha} \right) - \Psi \left( r_{ex}^{\alpha} \right)\right] \right\} }. \label{epsinv}
\end{eqnarray}
By substituting (\ref{epsinv}) into (\ref{TC_definition}),  the TC for a BPP is
\begin{eqnarray}
\tau_c\left(\zeta \right)
  =
\frac{ (1-\zeta) \left[ \log( 1 - \zeta ) + \beta_0 \Gamma^{-1} \right]  }{ A   \log \left\{ \kappa \left[ \Psi \left( r_{net}^{\alpha} \right) - \Psi \left( r_{ex}^{\alpha} \right)\right] \right\} }.
 \label{TC_BPP}
\end{eqnarray}

\begin{figure}[t]
\centering
\hspace{-0.5cm}
\includegraphics[width=9.25cm]{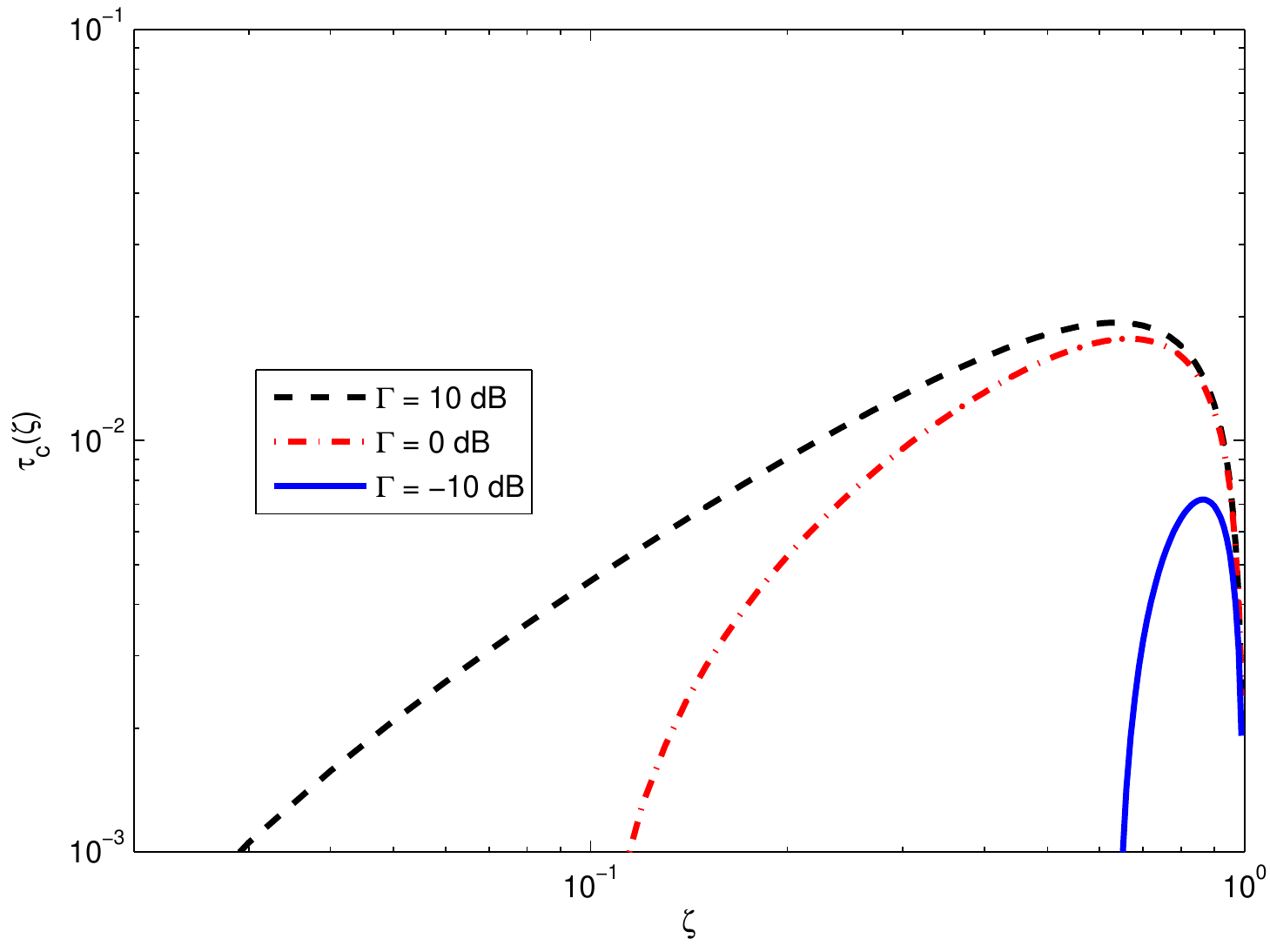}
\vspace{-0.5cm}
\caption{Transmission capacity $\tau_c(\zeta)$ as a function of the outage constraint $\zeta$ for different values of SNR $\Gamma$, when interferers are drawn from a BPP.  The network dimensions are $r_{ex}=0$ and $r_{net} = 2$.  The curves were produced using parameters $\beta=-10$ dB, $L'=1$, and $\alpha = 3$. Top curve: $\Gamma=10$ dB. Middle curve: $\Gamma=0$ dB.  Bottom curve: $\Gamma=-10$ dB.\label{Figure:Example6} }
\vspace{-0.5cm}
\end{figure}

{\bf Example \#6:}  In this example, a circular network is assumed with $r_{ex}=0$, $r_{net}=2$, and interferers drawn from a BPP.  The path-loss exponent is $\alpha =3$, and the equivalent number of frequency channels is $L'=1$, which results in $p=1$.  The SINR threshold is set to $\beta=-10$ dB.  Fig.
\ref{Figure:Example6} shows the transmission capacity as a function of the outage constraint $\zeta$ for three values of SNR
$\Gamma$. The curves were produced by using (\ref{TC_BPP}) and show that transmission capacity increases with $\Gamma$.

When the interferers are drawn from a PPP,  $\epsilon^{-1}(\zeta)$ is found by using solving $\epsilon =  F_{\mathsf Z}( \Gamma^{-1} ) = \zeta$ for $\lambda$, where $F_{\mathsf Z}( z )$ is given by (\ref{PPPNN}),
%
%
\begin{eqnarray}
\zeta
= 1 - \exp \left\{- \beta_0 \Gamma^{-1} -\pi \lambda \kappa^{-1} \left\{ 1- \kappa \left[ \Psi \left( r_{net}^{\alpha} \right) - \Psi \left( r_{ex}^{\alpha} \right)\right] \right\} \right \}. \nonumber
\end{eqnarray}
Solving for $\lambda$ and setting the result to $\epsilon^{-1}(\zeta)$,
\begin{eqnarray}
\epsilon^{-1}(\zeta)
 =
\frac{\log(1- \zeta)^{-1} - \beta_0 \Gamma^{-1}} {\pi  \kappa^{-1} \left\{ 1- \kappa \left[ \Psi \left( r_{net}^{\alpha} \right) - \Psi \left( r_{ex}^{\alpha} \right)\right]  \right\} }. \label{epsinv2}
\end{eqnarray}
By substituting (\ref{epsinv2}) into (\ref{TC_definition}),  the TC for a PPP is
\begin{eqnarray}
\tau_c\left(\zeta \right)
 =
 \frac{\left(1-\zeta \right) \left[ \log\left(1-\zeta \right)^{-1}- \beta_0 \Gamma^{-1}\right]}{\pi \kappa^{-1} \left\{1- \kappa \left[ \Psi \left( r_{net}^{\alpha} \right) - \Psi \left( r_{ex}^{\alpha} \right)\right]  \right\} }.
\label{tcppp}
\end{eqnarray}
When $r_{net} \rightarrow  \infty$, $r_{ex} = 0$ and $p=1$, (\ref{tcppp}) becomes
\begin{eqnarray}
\tau_c \left(\zeta \right)
 = \frac{\left( 1- \zeta\right) \left[\log \left( 1-\zeta \right)^{-1} - \beta_0 \Gamma^{-1} \right] }{\pi \beta_0^{\frac{2}{\alpha}}\frac{2 \pi}{\alpha} \csc \left( \frac{2 \pi}{\alpha}\right)}.
\label{Baccelli2}
\end{eqnarray}
This expression agrees with equation (4.10) in \cite{TransCap:2012}, which traces back to equation (62) of \cite{weber:2010} in the absence of noise.

\begin{figure}[t]
\centering
\hspace{-0.5cm}
\includegraphics[width=9.25cm]{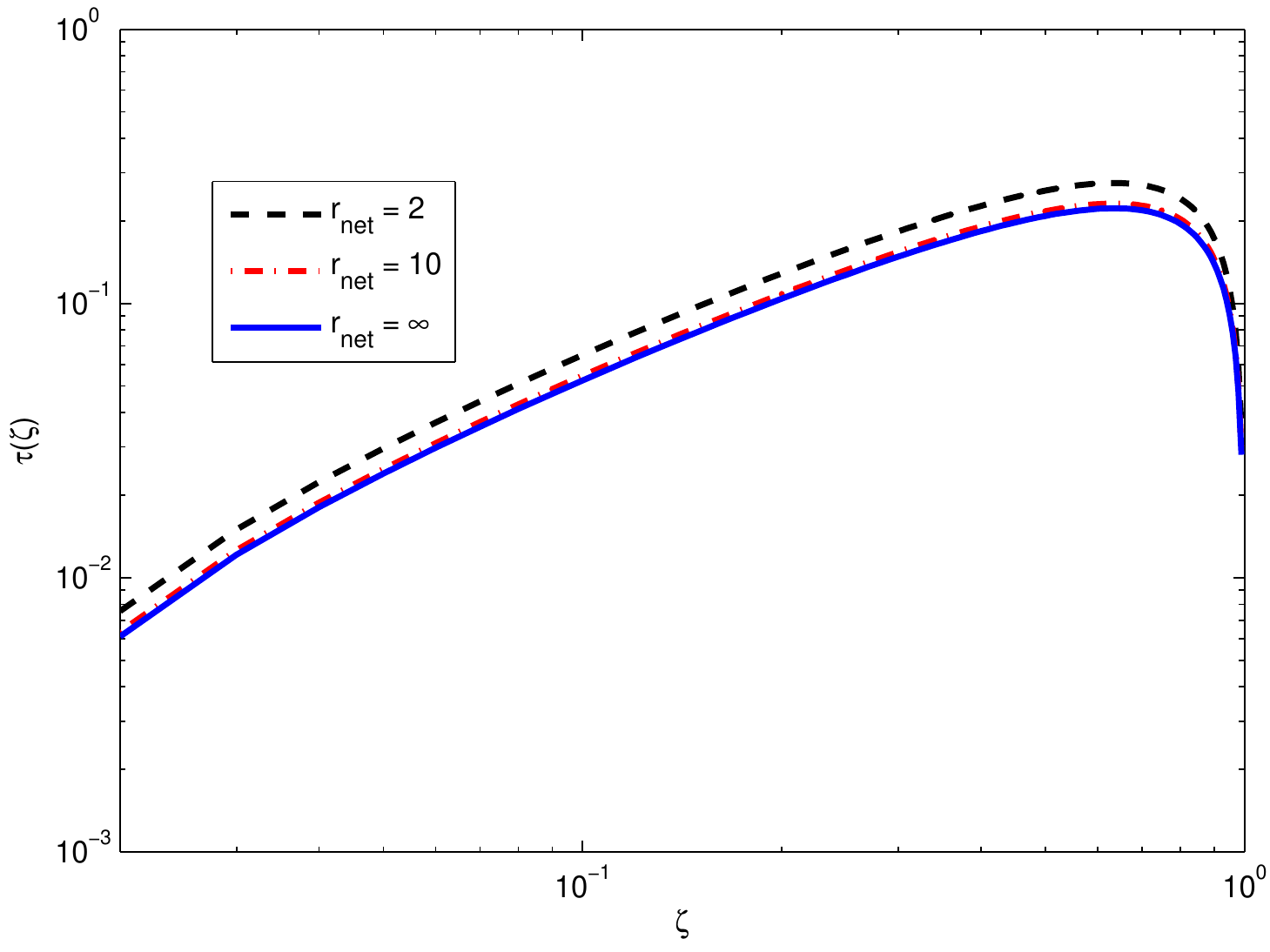}
\vspace{-0.5cm}
\caption{ Transmission capacity $\tau_c(\zeta)$ as a function of outage constraint $\zeta$ for three network radii, when interferers are drawn from a PPP and $r_{ex}=0$.   The curves were produced using parameters $\beta = -10$ dB , $\Gamma =10$ dB and $L'=1$. Top curve: $r_{net}=2$.  Middle curve: $r_{net}=10$. Bottom curve: $r_{net}=\infty$.
\label{Figure:Example7} }
\vspace{-0.5cm}
\end{figure}

{\bf Example \#7:}  In this example, a circular network is assumed with $r_{ex}=0$ and interferers drawn from a PPP.  Three different values of $r_{net}$ are considered: $r_{net} = \{2,10,\infty\}$.  The path-loss exponent is $\alpha =3$, and the equivalent number of frequency channels is $L'=1$, which results in $p=1$.  The SINR threshold is set to $\beta=-10$ dB, and the SNR is $\Gamma = 10$ dB.     Fig. \ref{Figure:Example7} shows the transmission capacity as a function of the outage constraint $\zeta$ for the three values of network radius $r_{net}$. The transmission capacity of the two networks with finite radius was computed using (\ref{tcppp}), while the transmission capacity of the infinite network was computed using ($\ref{Baccelli2}$). The curves show that the TC decreases with increasing network radius.

\section{Modulation-Constrained TC}
The transmission capacity expressions presented in the previous section are functions of the SINR threshold $\beta$ and make no assumptions about the existence of any particular type of modulation or channel coding.  In practice, the SINR threshold is a function of the modulation and coding that is used.   Let $C( \gamma )$ be the maximum achievable rate that can be supported by the chosen modulation at an instantaneous SINR of $\gamma$.  If a rate $R$ code is used, then an outage will occur when $C(\gamma) \leq R$.   Since $C(\gamma)$ is monotonic, it follows that $\beta$ is the value for which $C(\beta)=R$, and therefore we can write $\beta = C^{-1}(R)$.

Frequency-hopping systems often use noncoherent CPFSK modulation \cite{cheng:ciss2007,torrieri:2011}.  The maximum achievable rate of noncoherent CPFSK is given in \cite{cheng:ciss2007} for various modulation indices $h$, where it is called the {\em symmetric information rate}. In particular, Fig. 1 of \cite{cheng:ciss2007} shows the symmetric information rate of binary CPFSK as a function of $\gamma$ for various $h$.  To emphasize the dependence of the capacity on $h$, we use $C(h,\gamma)$ in the sequel to denote the rate of CPFSK with modulation index $h$.  For any value of $h$, the value of the SINR threshold $\beta$ can be found from the corresponding curve by finding the value of $\gamma$ for which $C(h,\gamma)=R$.  For instance, when $R=1/2$ and $h=1$, the required $\beta = 3.7$ dB.  In \cite{torrieri:2008}, it was found that in practice, and over a wide range of code rates, turbo-coded noncoherent CPFSK is consistently about 1 dB away from the corresponding modulation-constrained capacity limit.  Thus, the $\beta$ required in practice will generally be higher than the value of $\beta_{min}(R,h)$ by a small margin.  For instance, if a 1 dB margin is used, then the SINR threshold for noncoherent binary CPFSK with $R=1/2$ and $h=1$ should be set to $\beta = 4.7$ dB.

When accounting for modulation and coding, the maximum data transmission rate is determined by the bandwidth $B/L$ of a frequency channel, the spectral efficiency of the modulation, and the code rate.  Let $\eta$ be the spectral efficiency of the modulation, given in symbols per second per Hz, and defined by the symbol rate divided by the 99 percent-power bandwidth of the modulation\footnote[1]{Percent-power bandwidths other than 99 can be used, but will influence the amount of adjacent-channel interference.}.  The spectral efficiency of CPFSK can be found by numerically integrating the normalized power-spectral densities given in \cite{torrieri:2011}, or since we assume many symbols per hop, by Equation (3.4-61) of \cite{proakis:2008} and then inverting the result.  To emphasize the dependence of $\eta$ on $h$, we denote the spectral efficiency of CPFSK as $\eta(h)$ in the sequel.  When combined with a rate-$R$ code, the spectral efficiency of CPFSK becomes $R \eta(h)$ (information) bits per second per Hz, where $R$ is the ratio of information bits to code symbols.  The data rate supported by the channel is $R \eta(h) B/L$ bits per second.  The average data rate, or throughput, must account for the duty factor $d$ and only count correct transmissions.  Hence, the throughput is
\begin{eqnarray}
   T
   & = &
   \frac {   R \eta(h) B d (1-\epsilon) }{L}
   =
   \frac {   R \eta(h) B (1-\epsilon) }{L'}.
\end{eqnarray}

The {\em modulation-constrained} transmission capacity is the throughput multiplied by the node density,
\begin{eqnarray}
   \tau (\lambda)
   & = &
   \mathcal \lambda T = \frac{\lambda R \eta(h) B (1-\epsilon) }{L'}.\label{Equation:TC}
\end{eqnarray}
In contrast with (\ref{TC_definition}), this form of transmission capacity explicitly takes into account the code rate $R$, as well as the spectral efficiency of the modulation $\eta(h)$.  It furthermore accounts for the hopping bandwidth $B/L'$.  Rather than constraining outage probability, $\tau( \lambda )$ fixes the node density $\lambda$ and allows the outage probability to vary accordingly.  Since it accounts for the actual system bandwidth $B$, (\ref{Equation:TC}) assumes units of $bps/m^2$.  By dividing by bandwidth, the {\em normalized} modulation-constrained transmission capacity
\begin{eqnarray}
  \tau'(\lambda)
  & = &
  \frac{\tau}{B}
  =
  \frac{\lambda R \eta(h) (1-\epsilon) }{L'}.\label{Equation:TCnorm}
\end{eqnarray}
takes on units of $bps/Hz/m^2$.  However, unlike (\ref{TC_definition}), $\tau'(\lambda)$ is in terms of {\em information} bits rather than {\em channel} bits.


\section{Network Optimization \\ by an Exhaustive search }\label{Section:Optimization}

The main goal of this paper is to find the $(L',R,h)$ that maximizes the normalized TC $\tau'(\lambda)$ for a frequency-hopping ad hoc network, assuming that transmissions occur using a capacity-approaching code (e.g., turbo or LDPC) and noncoherent binary continuous-phase frequency shift keying (CPFSK) modulation. The optimization can be accomplished using an exhaustive search by performing the following steps:
\begin{enumerate}
   \item \label{pickBetaE} Pick a value of $\beta$.
   \item \label{pickhE} Pick a value of $h$, and determine the rate $R$ corresponding to the current $\beta$ (this is found by setting $R=C(h,\beta)$) and its corresponding bandwidth efficiency $ \eta(h)$.

   \item \label{pickLE} Pick a value of $L'$.
   \item Use (\ref{cdf_BPP}) to compute the average outage probability $\epsilon_{M}$ if the interferers are drawn from a BPP or (\ref{cdf_PPP}) to compute $\epsilon$ if the interferers are drawn from a PPP . \label{OPESE}
   \item For the set of $(h,R)$ found in step \ref{pickhE}, determine $\tau'(\lambda)$ by using (\ref{Equation:TCnorm}).
   \item Return to step \ref{pickLE} until all $L'$ are considered.
   \item Return to step \ref{pickhE} until all $h$ are considered.
   \item Return to step \ref{pickBetaE} until all $\beta$ are considered.

\end{enumerate}

The above procedure will find the $\tau'(\lambda)$ for each $(L',R,h)$ considered, and the optimal value of these parameters are the ones that maximize $\tau'(\lambda)$.
By limiting $L'$ to be integer valued (which is not necessary if $d$ is a fraction), the number of values is finite and an exhaustive search up to some maximum value is feasible.  The value of $\beta$ is continuous, and therefore must be quantized.  For the exhaustive search results presented in this section, $\beta$ was quantized to a spacing of $0.1$ dB over the range $-2$ dB $\leq \beta \leq 12$ dB, and $h$ was quantized to a spacing of 0.01 over the range $0 \leq h \leq 1$.


\subsection{Optimization Results for a BPP}\label{Section:Results_BPP}

\begin{figure}[t]
\centering
\hspace{-0.5cm}
\includegraphics[width=9.25cm]{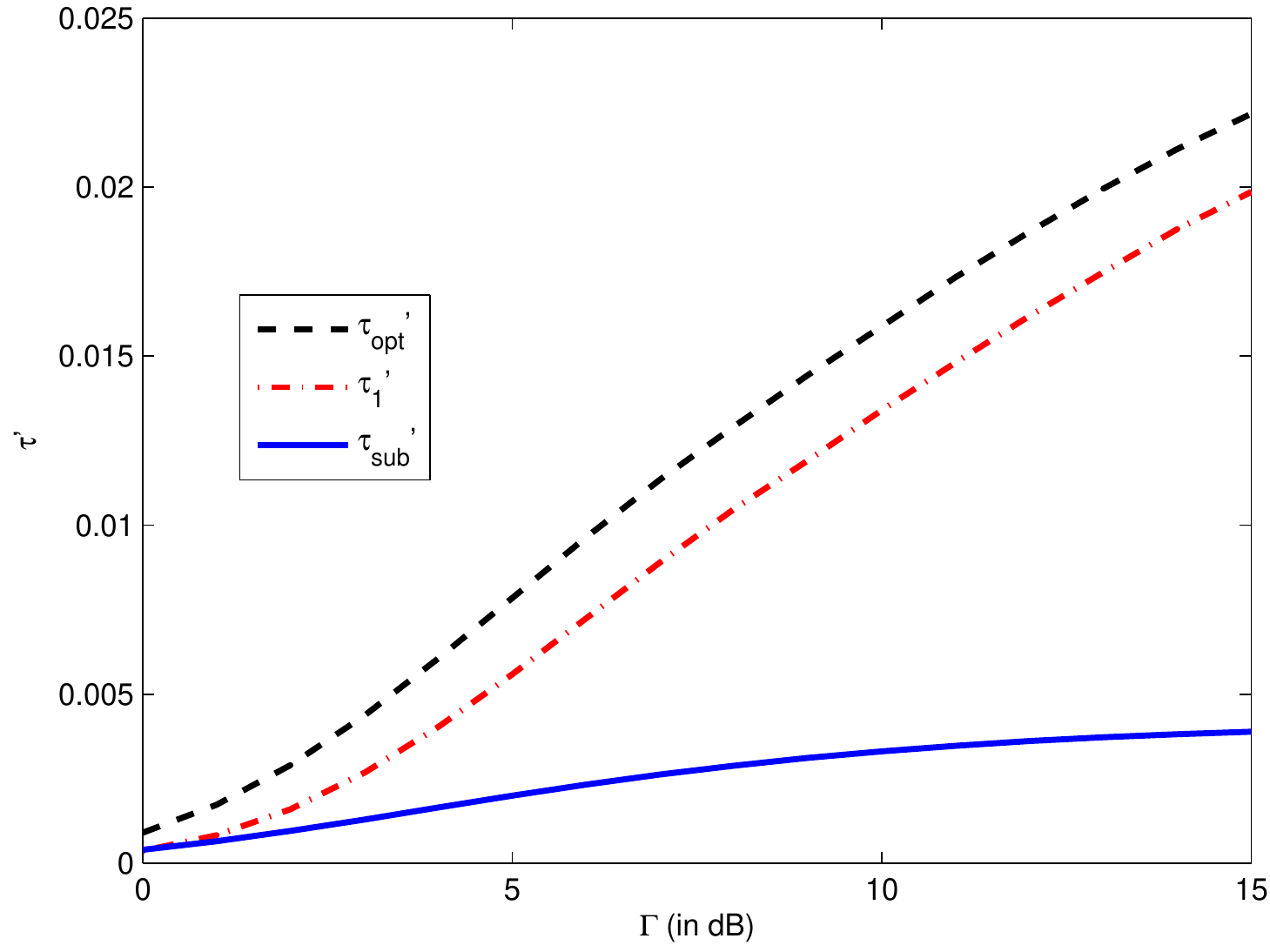}
\vspace{-0.5cm}
\caption{Maximum normalized modulation-constrained TC $\tau'(\lambda)$ as a function of the SNR $\Gamma$ for three cases: (1) the code achieves capacity; (2) the  code has a 1 dB gap from capacity; and (3) a typical set of ($(L',R,h)$)=($200,1/2,1$) is used.  The interferers are drawn from a BPP.
     \label{Figure:Example8} }
\end{figure}

The optimization was run for a network of $M=50$ interferers placed according to a BPP with an inner radius of $r_{ex} = 0.25$ and an outer radius $r_{net} = 2$.  The path-loss exponent was fixed to $\alpha=3$.
Fig. \ref{Figure:Example8} shows the maximum normalized modulation-constrained TC $\tau_{opt}'(\lambda)$ as function the SNR $\Gamma$.  For each value of $\Gamma$, the optimal set of $(L',R,h)$ that maximizes the TC was found using the previously described exhaustive search.

The value of $\tau_{opt}'(\lambda)$ was computed assuming a capacity-achieving code.  Suppose that instead, the code has a {\em gap} of 1 dB from capacity, i.e. that the required threshold $\beta$ is 1 dB higher than that predicted by information theory.  The transmission capacity will be lower due to this gap.  The curve labeled $\tau_{1}'$ shows the TC of a code when using a code with a 1 dB gap from capacity when using the optimal values of $(L',R,h)$ found assuming a capacity-achieving code.  As can be seen, a modest loss in TC occurs when the code has a 1 dB gap from capacity. In addition, Fig. \ref{Figure:Example8} also shows the normalized TC $\tau_{sub}'(\lambda)$ of a system with a suboptimal but typical choice of parameters: $(L',R,h) = (200,1/2,1)$.  The results shown in Fig. \ref{Figure:Example8} highlight the importance of parameter optimization.  The TC is improved by a factor of 5-10 by selecting optimal, rather than arbitrary parameters.

\begin{figure}[t]
\centering
\hspace{-0.5cm}
\includegraphics[width=9.25cm]{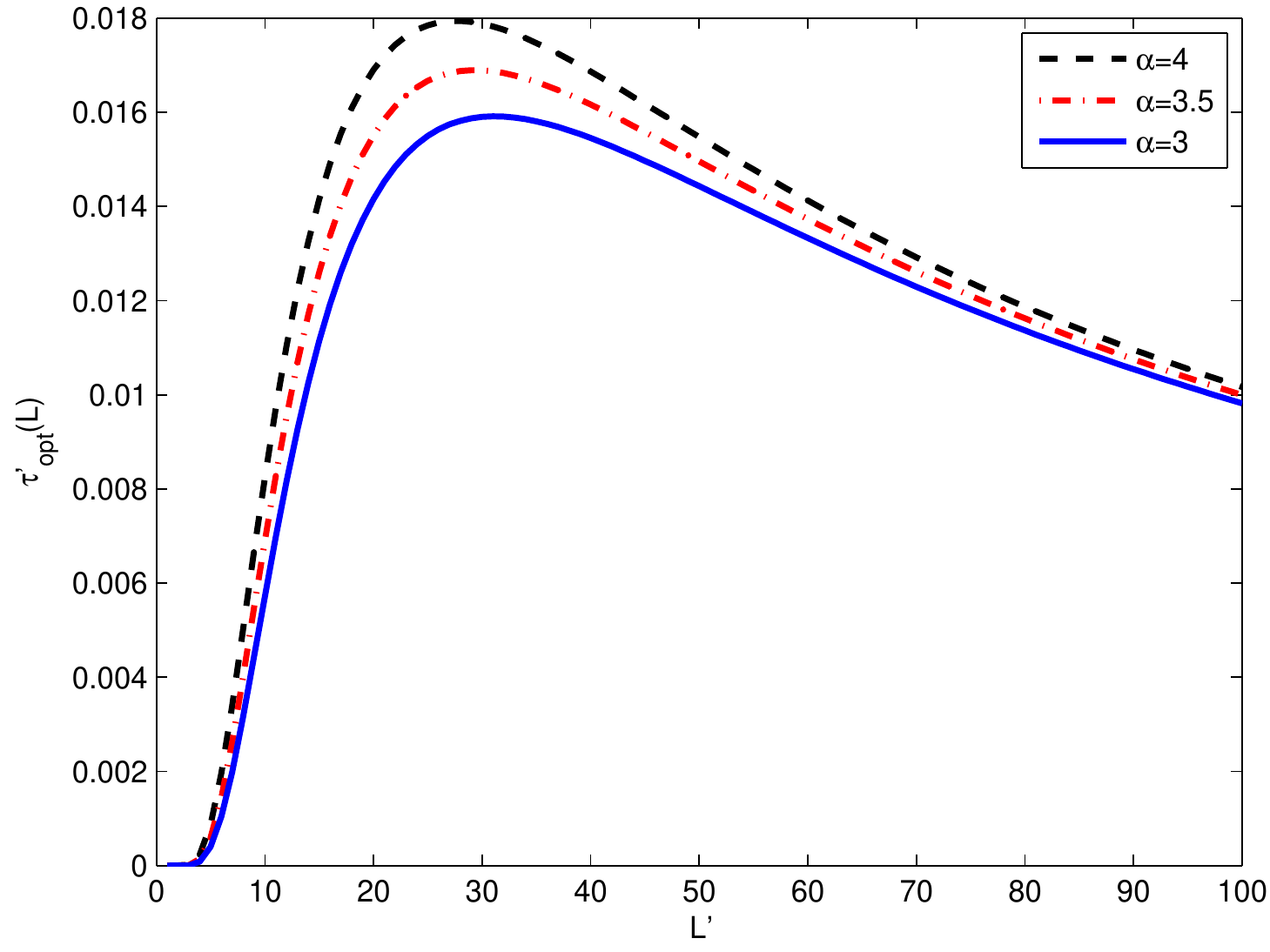}
\vspace{-0.5cm}
\caption{Maximum normalized modulation-constrained TC $\tau_{opt}'(\lambda)$ as a function of the equivalent number of frequency channels $L'$. The interferers are drawn from a BPP and the network dimensions are  $r_{ex}$ = 0.25 and $r_{net}$ = 2. For each value of $L'$, the optimal $R$ and $h$ are found. Top curve: $\alpha$ = 4.  Middle curve: $\alpha$ = 3.5.  Bottom curve: $\alpha$ = 3.
\label{Figure:Example9} }
\hspace{-0.5cm}
\end{figure}

\begin{figure}[t]
\centering
\hspace{-0.5cm}
\includegraphics[width=9.25cm]{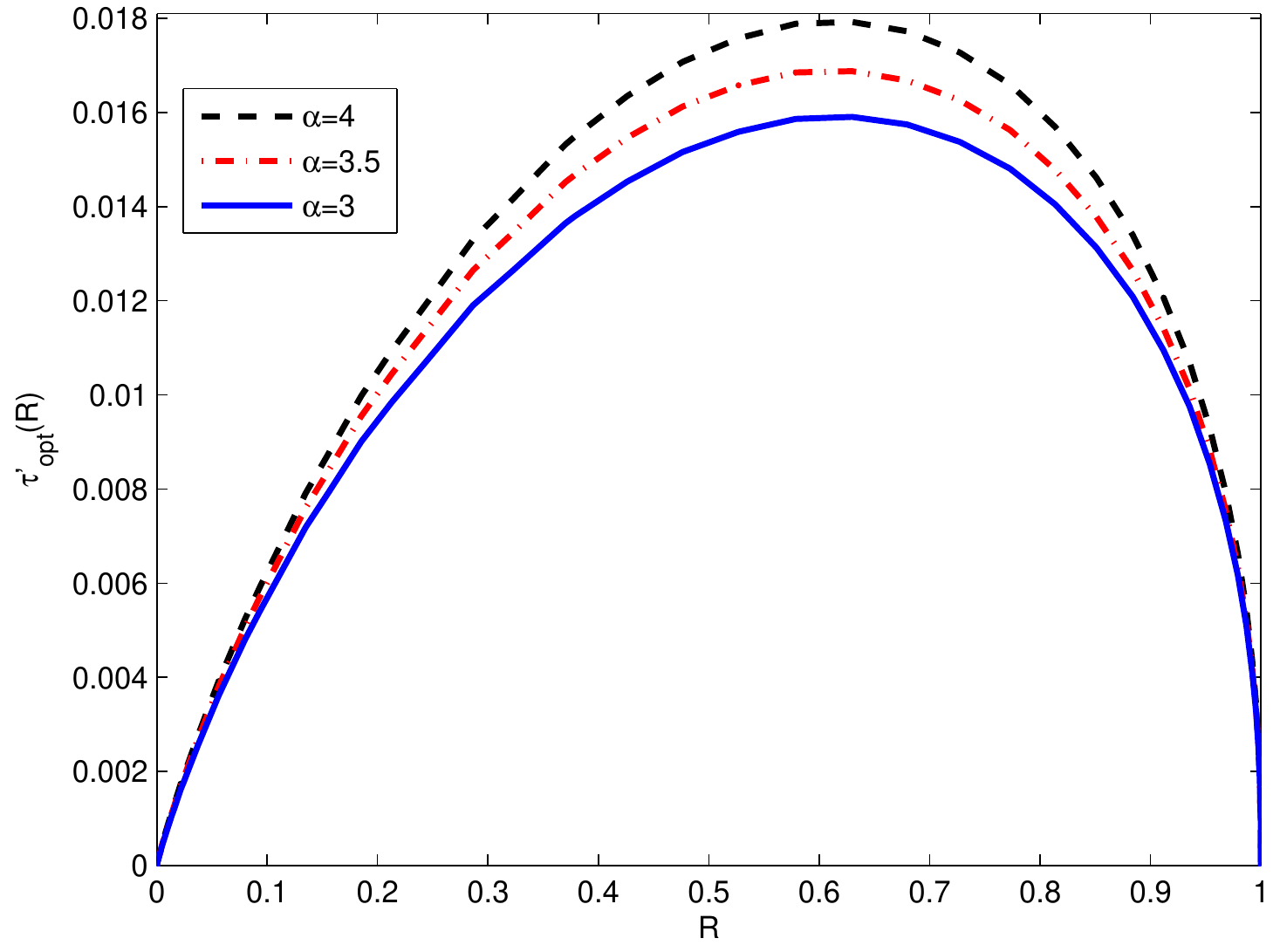}
\vspace{-0.5cm}
\caption{Maximum normalized modulation-constrained TC $\tau_{opt}'(\lambda)$ as a function of the code rate $R$. The interferers are drawn from a BPP and the network dimensions are  $r_{ex}$ = 0.25 and $r_{net}$ = 2. For each value of $R$, the optimal $L'$ and $h$ are found. Top curve: $\alpha$ = 4.  Middle curve: $\alpha$ = 3.5.  Bottom curve: $\alpha$ = 3. \label{Figure:Example10} }
\end{figure}

\begin{figure}[t]
\centering
\hspace{-0.5cm}
\includegraphics[width=9.25cm]{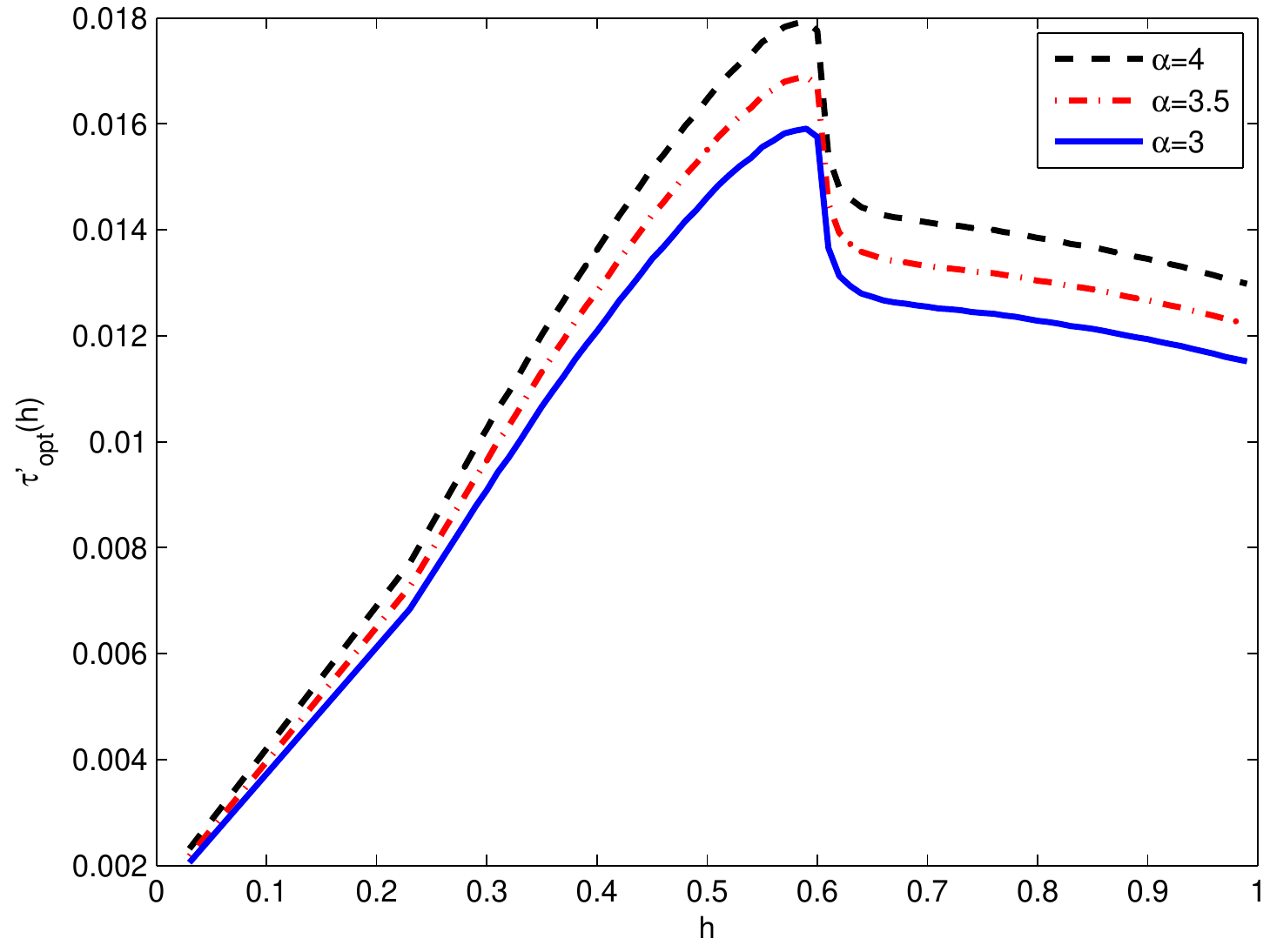}
\vspace{-0.5cm}
\caption{Maximum normalized modulation-constrained TC $\tau_{opt}'(\lambda)$ as a function of the modulation index $h$. The interferers are drawn from a BPP and the network dimensions are  $r_{ex}$ = 0.25 and $r_{net}$ = 2. For each value of $h$, the optimal $L'$ and $R$ are found. Top curve: $\alpha$ = 4.  Middle curve: $\alpha$ = 3.5.  Bottom curve: $\alpha$ = 3.\label{Figure:Example11} }
\vspace{-0.5cm}
\end{figure}

Fig. \ref{Figure:Example9}-\ref{Figure:Example11} explore the relative importance of each of the three parameters.  In each curve, the SNR was set to $\Gamma = 10$ dB and one parameter is varied. At each value of the parameter, the TC is maximized with respect to the other two parameters.  Three values of path-loss exponent are considered, $\alpha = \{3,3.5,4\}$.  The optimal values of each of the parameters can be identified by locating the peaks of each curve.  A general trend is that the TC improves with increasing $\alpha$, though the optimal parameter values are not strongly influenced by the $\alpha$.

\subsection{Optimization Results for a PPP}\label{Section:Results_PPP}
\begin{figure}[t]
\centering
\hspace{-0.5cm}
\includegraphics[width=9.25cm]{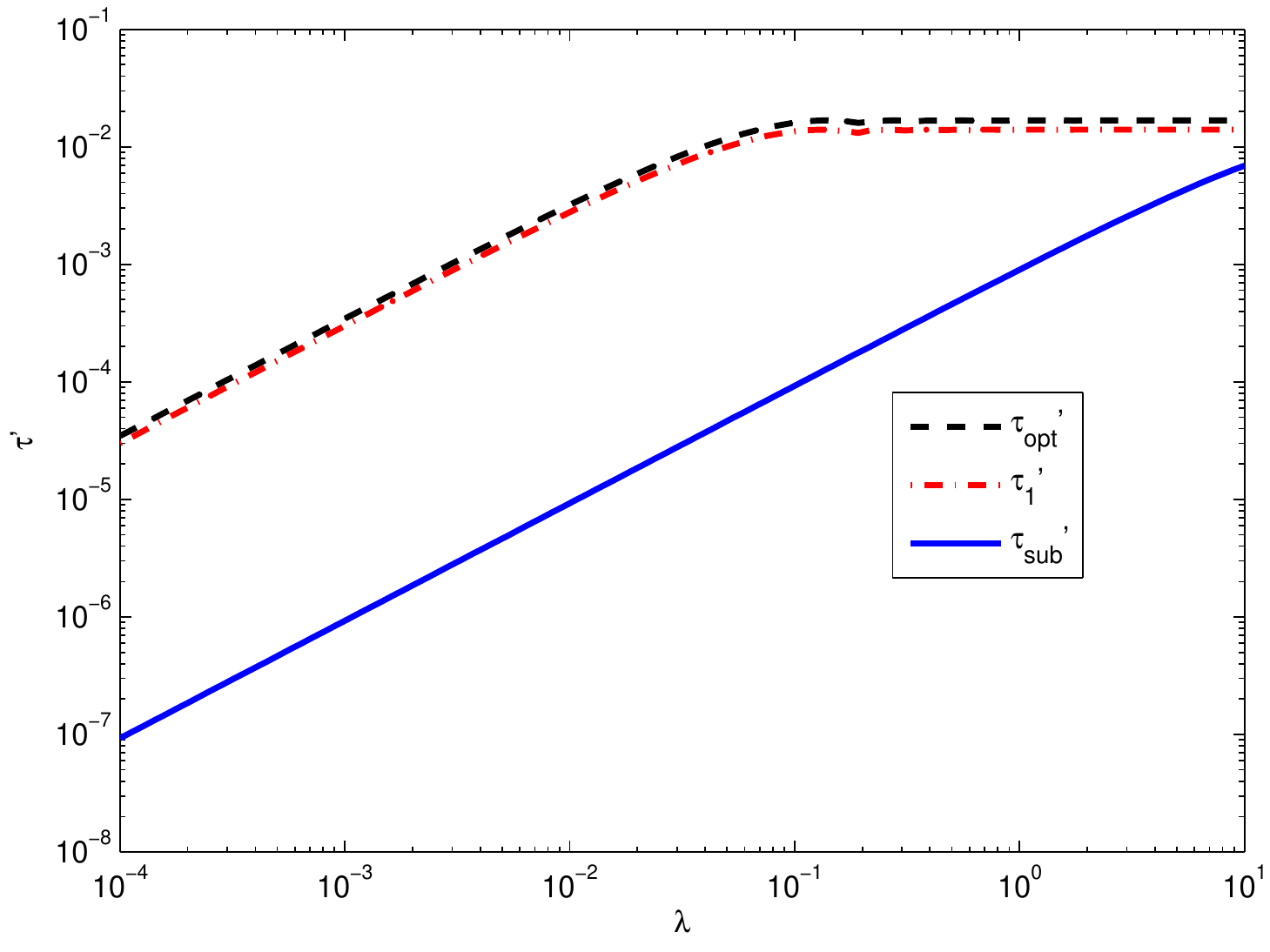}
\vspace{-0.5cm}
\caption{Maximum normalized modulation-constrained TC $\tau'(\lambda)$  as function of mobile density $\lambda$ for three cases: (1) the code achieves capacity; (2) the  code has a 1 dB gap from capacity; and (3) a typical set of ($(L',R,h)$)=($200,1/2,1$) is used.  The interferers are drawn from a PPP. \label{Figure:Example12} }
\end{figure}

Next, the optimization was run for a network with interferers drawn from a PPP with an inner radius of $r_{ex} = 0.25$ and an outer radius $r_{net} = 2$.
The SNR was set to $\Gamma=10$ dB and the path-loss exponent to $\alpha=3$.   Fig. \ref{Figure:Example12} shows the maximum modulation-constrained normalized TC $\tau_{opt}'(\lambda)$ as function  of the mobile density $\lambda$.  For per each value of $\lambda$, the optimal set of $(L',R,h)$ that maximizes the TC was found using the previously described exhaustive search. Similar to Fig. \ref{Figure:Example8}, Fig. \ref{Figure:Example12} shows the performance $\tau'_1(\lambda)$ when the code has a 1 dB gap from capacity, and shows the performance $\tau_{sub}'(\lambda)$ of a system that uses the typical choice of parameters: $(L',R,h) = (200,1/2,1)$.  While the loss due to using a code with a 1 dB gap from capacity is quite minimal, the loss due to using suboptimal parameters is quite high, especially in sparser networks.

\begin{figure}[t]
\centering
\hspace{-0.5cm}
\includegraphics[width=9.25cm]{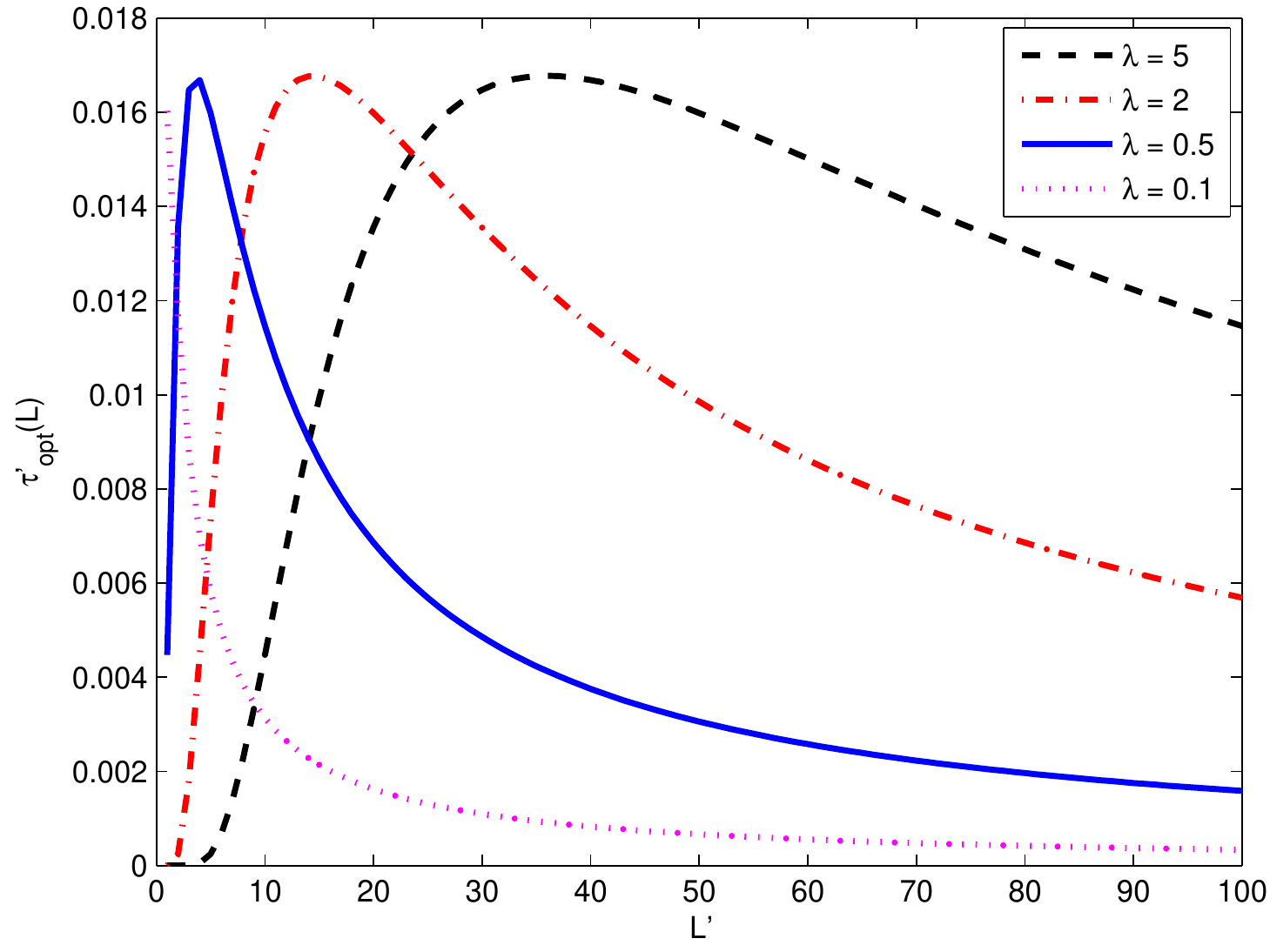}
\vspace{-0.5cm}
\caption{Maximum normalized modulation-constrained TC $\tau_{opt}'(\lambda)$ as a function of the equivalent number of frequency channels $L'$. The interferers are drawn from a PPP and the network dimensions are  $r_{ex}$ = 0.25 and $r_{net}$ = 2. For each value of $L'$, the optimal $R$ and $h$ are found.
Curves from top to bottom: (1) $\lambda$ = 5;  (2) $\lambda$ = 2; (3) $\lambda$ = 0.5; (4) $\lambda$ = 0.1.
\label{Figure:Example13} }
\end{figure}

\begin{figure}[t]
\centering
\hspace{-0.5cm}
\includegraphics[width=9.25cm]{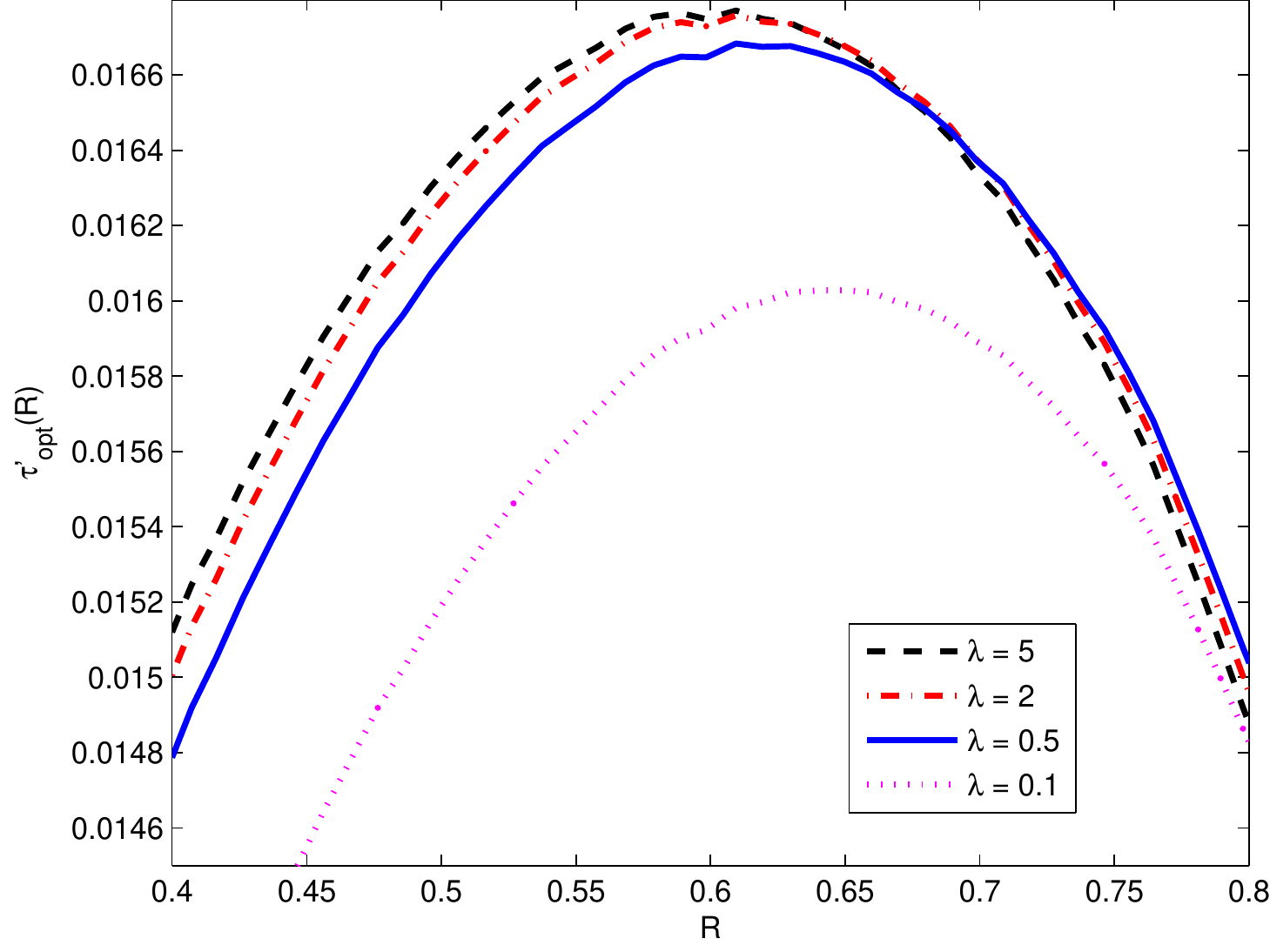}
\vspace{-0.5cm}
\caption{Maximum normalized modulation-constrained TC $\tau_{opt}'(\lambda)$ as a function of the code rate $R$. The interferers are drawn from a PPP and the network dimensions are  $r_{ex}$ = 0.25 and $r_{net}$ = 2. For each value of $R$, the optimal $L'$ and $h$ are found.
Curves from top to bottom: (1) $\lambda$ = 5;  (2) $\lambda$ = 2; (3) $\lambda$ = 0.5; (4) $\lambda$ = 0.1.
(4) $\lambda$ = 0.1. \label{Figure:Example14} }
\end{figure}

\begin{figure}[t]
\hspace{-0.5cm}
\centering
\hspace{-0.5cm}
\includegraphics[width=9.25cm]{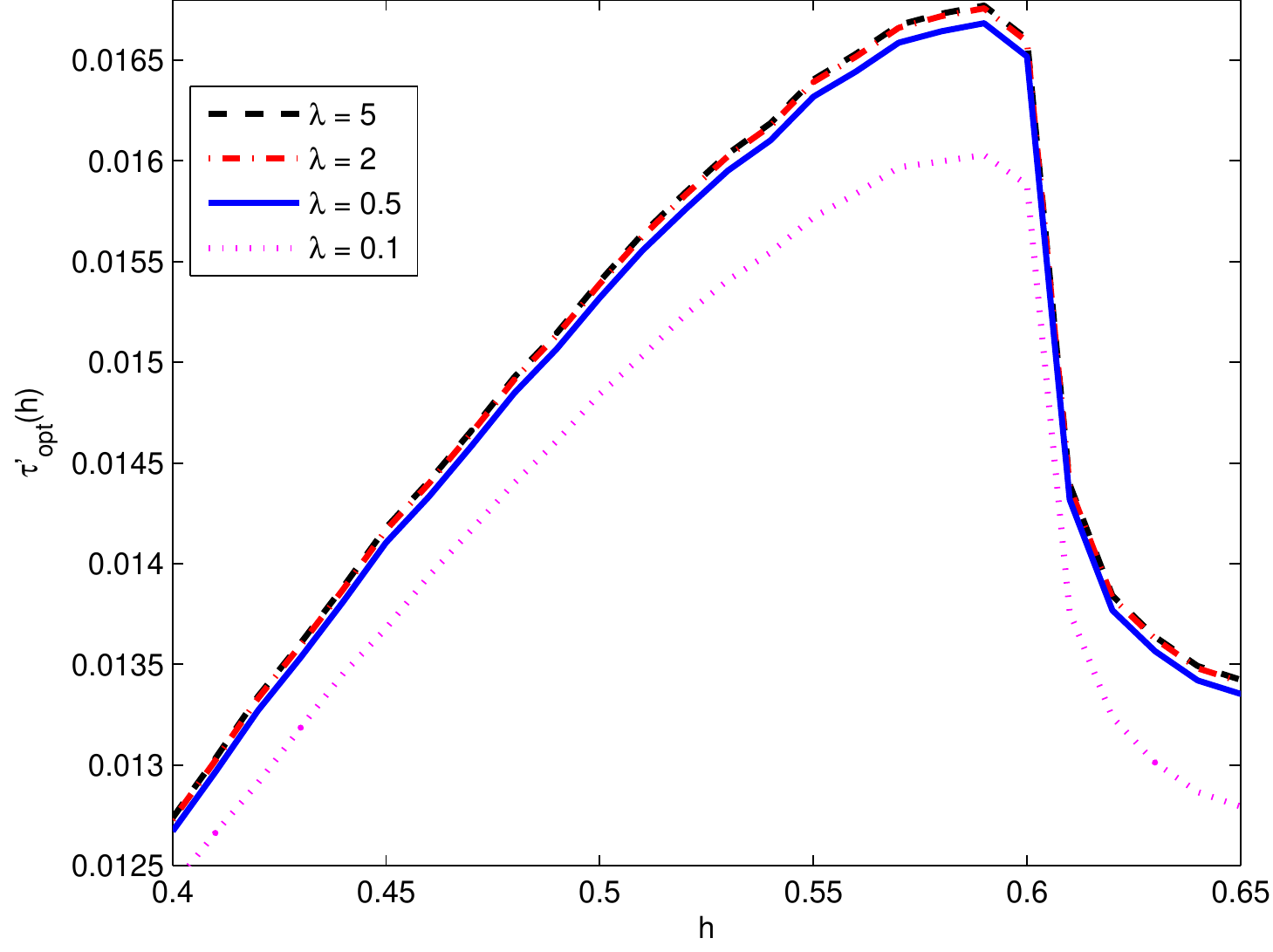}
\vspace{-0.5 cm}
\caption{Maximum normalized modulation-constrained TC $\tau_{opt}'(\lambda)$ as a function of the modulation index $h$. The interferers are drawn from a PPP and the network dimensions are  $r_{ex}$ = 0.25 and $r_{net}$ = 2. For each value of $h$, the optimal $L'$ and $R$ are found.
Curves from top to bottom: (1) $\lambda$ = 5;  (2) $\lambda$ = 2; (3) $\lambda$ = 0.5; (4) $\lambda$ = 0.1. \label{Figure:Example15} }
\end{figure}

Fig. \ref{Figure:Example13}-\ref{Figure:Example15} explore the relative importance of each of the three parameters by varying one parameter.  At each value of the parameter, the modulation-constrained TC is maximized with respect to the other two parameters.  Four values of mobile density are considered, $\lambda = \{0.1, 0.5, 2, 5\}$.  The optimal values of each of the parameters can be identified by locating the peaks of each curve.

In Fig. \ref{Figure:Example13}, it is shown the maximum normalized TC as a function of the equivalent number of frequency channels $L'$ for a certain annular network area that has three different density of interferers per unit Area ($\lambda=5$, $\lambda=2$ and $\lambda=0.5$ ) when for each of them the optimal modulation index and the optimal rate of the coded modulation are used.   Fig. \ref{Figure:Example13} shows a strong dependence on $L'$, with the optimal $L'$ becoming larger with increasing network density.  Fig. \ref{Figure:Example14} shows a relatively weak dependence on $R$, with denser networks requiring a slightly lower rate. Fig. \ref{Figure:Example15} shows a weak dependence on $h$, with $h \approx 0.59$ providing optimal performance for every density.

\section{Network Optimization by Gradient search method}\label{Section:Optimization_GM}
\begin{table}
  \centering
  \caption{Results of the Optimization for an annular network area where the interferers are drawn from a BPP. The number of interferers is fixed to $M=50$.
  \label{maintable}}
   \vspace{-0.3cm}
  \begin{tabular}{|c|c|c|c|c|c|c|c|c|}
  \hline
  $r_{net}$ & $r_{ex}$ & $\alpha$  & $\tau'_{opt} $ & $\tau'_{opt_{\nabla}} $ & $I_{\nabla}$ \\
  \hline
  1         & 0.25            &  3       &  0.04427  & 0.04427  & 66  \\
  \cline{3-6}
            &              &  3.5     &  0.04395 & 0.04395 & 66  \\
  \cline{3-6}
            &              &  4       & 0.04372 & 0.04372 & 66  \\
  \cline{2-6}
            & 0.5            &  3       & 0.04503 & 0.04503 & 87  \\
  \cline{3-6}
            &              &  3.5      & 0.04468 & 0.04468 & 66   \\
  \cline{3-6}
            &              &  4       &  0.04437 & 0.04437 & 66   \\
  \hline
  2         & 0.25            &  3      &  0.01590  & 0.01590 & 54   \\
  \cline{3-6}
            &              &  3.5       &  0.01688  & 0.01688 & 54 \\
  \cline{3-6}
            &              &  4       &    0.01792  & 0.01792 & 54    \\
  \cline{2-6}
            & 0.5            &  3      &     0.01641  & 0.01641 & 63  \\
  \cline{3-6}
            &              &  3.5      &   0.01752  & 0.01752 & 63   \\
  \cline{3-6}
            &              &  4       &    0.01871  & 0.01871 & 57  \\
  \hline
  4        & 0.25            &  3       &  0.00983  & 0.00983 & 50  \\
  \cline{3-6}
            &              &  3.5       &  0.01187  & 0.01187 & 47  \\
  \cline{3-6}
            &              &  4      &   0.01395  & 0.01395 & 50 \\
  \cline{2-6}
            & 0.5            &  3       &    0.01024  & 0.01024 & 44   \\
  \cline{3-6}
            &              &  3.5      &   0.01252  & 0.01252 & 44   \\
  \cline{3-6}
            &              &  4      &    0.01484 & 0.01484 & 47  \\
  \hline
  \end{tabular}
\end{table}
\begin{table}
  \centering
  \caption{Results of the Optimization for an annular network area where the interferers are drawn from a PPP. The intensity $\lambda$ per unit area is fixed to $\lambda=1$. \label{maintable1}}
     \vspace{-0.3cm}
  \begin{tabular}{|c|c|c|c|c|c|c|c|c|}
  \hline
  $r_{net}$ & $r_{ex}$ & $\alpha$  & $\tau'_{opt} $ & $\tau'_{opt_{\nabla}} $ & $I_{\nabla}$ \\
  \hline
  1         & 0.25            &  3       &  0.04654   & 0.04654  & 129   \\
  \cline{3-6}
            &              &  3.5     & 0.04623 & 0.04623 & 129  \\
  \cline{3-6}
            &              &  4       & 0.04598 & 0.04598 &  129  \\
  \cline{2-6}
            & 0.5            &  3       & 0.05932 &  0.05932 & 129 \\
  \cline{3-6}
            &              &  3.5      & 0.05881 & 0.05881 & 129   \\
  \cline{3-6}
            &              &  4       &  0.05838 & 0.05838 & 129  \\
  \hline
  2         & 0.25            &  3      &   0.01597  & 0.01597 &  126    \\
  \cline{3-6}
            &              &  3.5       &  0.01697  & 0.01697 & 126   \\
  \cline{3-6}
            &              &  4       &   0.01801  & 0.01801 & 134   \\
  \cline{2-6}
            & 0.5            &  3      &     0.01731  &  0.01731 & 134   \\
  \cline{3-6}
            &              &  3.5      &   0.01845  & 0.01845 & 134  \\
  \cline{3-6}
            &              &  4       &    0.01973  & 0.01973 & 123 \\
  \hline
  4        & 0.25            &  3       &  0.00977 & 0.00977 & 117 \\
  \cline{3-6}
            &              &  3.5       &   0.01180  & 0.01180 & 117   \\
  \cline{3-6}
            &              &  4      &    0.01387  &  0.01387 & 117  \\
  \cline{2-6}
            & 0.5            &  3       &    0.01030  & 0.01030 &  120   \\
  \cline{3-6}
            &              &  3.5      &    0.01258  &  0.01258 & 135   \\
  \cline{3-6}
            &              &  4      &     0.01491  & 0.01491 &  123  \\
  \hline
  \end{tabular}
  \vspace{-0.5cm}
\end{table}
The results of the exhaustive search presented in the previous section suggest that the modulation-constrained TC is a concave function of ($L',R,h$).  It follows that the optimization is a convex optimization problem and can be efficiently solved through a gradient-search method \cite{boyd:2004}.  In particular, the optimization can be accomplished by performing the following steps:
\begin{enumerate}
   \item Pick intervals for $L'$ ($[L'_{min},L'_{max}]$), $\beta$ ($[\beta_{min},\beta_{max}]$) and $h$ ($[h_{min},h_{max}]$). \label{interval}
   \item Create sets $ L_{set}=\{ L'_{min}, \left(L'_{max} + L'_{min} / 2 \right), L'_{max}  \}$, $ \beta_{set}=\{ \beta_{min}, \left(\beta_{max} + \beta_{min} / 2 \right), \beta_{max}  \}$, and $ h_{set}=\{ h_{min}, \left(h_{max} + h_{min} / 2 \right), h_{max}  \}$     composed of the two extreme points and center point of each interval.
\label{vector}
   \item Pick one of the three values of $\beta$.
   \item Pick one of the three values of $h$ and determine the rate $R$ corresponding to the current $h$ and $\beta$ (this is found by setting $R=C(h,\beta)$) and its corresponding bandwidth efficiency $ \eta(h)$. \label{opth}
   \item For all three values of $L'$ and for the set of $(h,R)$ found in the last step, determine $\tau'(\lambda)$ by using (\ref{Equation:TCnorm}). \label{optL}
   \item Once $\tau'(\lambda)$ is computed for all three values of $L'$, determine which value has the largest normalized TC:
       \begin{enumerate}
       \item If the maximum is at one of the two external points, the center of the search points is moved in that direction and two new external points are chosen closer to the new center point;
       \item If the maximum is at the center point, the two external points are moved closer to the center. Each time the maximum is in the center, the distance between the two external points and the center is gradually decreased.
       \end{enumerate} \label{optL1}
   \item Return to step \ref{optL} and use the new three points, until the distance between the two external points and the center reaches the values fixed in step \ref{interval} and the maximum stays in the center. \label{stepL}
   \item Repeat step \ref{optL}, \ref{optL1} and \ref{stepL}, for all three values of $h$ and save the normalized TC of them when $L$ is optimized. \label{opth}
   \item As step \ref{optL1}, once $\tau'(\lambda)$ is computed for all three values of $h$, determine which value has the largest normalized TC:
       \begin{enumerate}
       \item If the maximum is at one of the two external points, the center of the search points is moved in that direction and two new external points are chosen closer to the new center point;
       \item If the maximum is at the center point, the two external points are moved closer to the center. Each time the maximum is in the center, the distance between the two external points and the center is gradually decreased.
       \end{enumerate} \label{opth1}
   \item Return to step \ref{opth} and use the new three points of $h$, until the distance between the two external points is sufficiently small and the maximum point remains in the center. \label{steph}
   \item Repeat step \ref{opth}, \ref{opth1} and \ref{steph}, for all three values of $\beta$ and save the normalized TC of each.    \label{optbeta}
   \item As step \ref{opth1}, once $\tau'(\lambda)$ is computed for all three values of $\beta'$, determine which value has the largest normalized TC:\label{optbeta1}
          \begin{enumerate}
       \item If the maximum is at one of the two external points, the center of the search points is moved in that direction and two new external points are chosen closer to the new center point;
       \item If the maximum is at the center point, the two external points are moved closer to the center. Each time the maximum is in the center, the distance between the two external points and the center is gradually decreased.
       \end{enumerate}
   \item Return to step \ref{optbeta} and use the new three points of $\beta$, until the distance between the two external points is sufficiently small and the maximum point remains in the center.
\end{enumerate}

The algorithm is initialized by the initial intervals selected at step 1.  As the algorithm runs, the size of the intervals get successively smaller.  The algorithm can stop once both of the following conditions are satisfied at the same time: (1) the optimal values of $L'$, $h$ and $\beta$ are the same of the previous iteration; (2) the difference between each of the three elements inside the sets $L_{set}$, $h_{set}$ and $\beta_{set}$ is equal to the quantization step for respectively $L$, $h$ and $\beta$ .

Tables \ref{maintable} and  \ref{maintable1} compare the results of optimizations performed by exhaustive search and gradient search for networks distributed according to BPP and PPP processes, respectively.  The column marked
$\tau_{opt}$ is the maximum modulation-constrained TC found by using the exhaustive-search technique of Section \ref{Section:Optimization}, while $\tau_{opt_{\nabla}}$ is the value found using the gradient-search technique presented in this section. For each type of spatial distribution, three values of $r_{net}$, two values of $r_{ex}$, and three values of $\alpha$ were considered.  For the BPP, the number of interferers was set to $M=50$, while for the PPP, the density was set to $\lambda = 1$.  For both processes, the SNR was set to $\Gamma = 10$ dB.

The tables also indicate the number of iterations required for the gradient-search technique to converge.  The value is indicated by the column marked $I_{\nabla}$.  Each iteration requires that 200 values of $\tau'(\lambda)$ be evaluated, since for L was used a spacing of 1 over the range $1 \leq L \leq 200$.  Notice the slight variation in the number of iterations. This is in contrast with the exhaustive-search algorithm, which requires that fixed number of values of $\tau'(\lambda)$ be evaluated.  In particular, the exhaustive-search optimization considered $2,848,200$ sets of discretized parameters by using the same parameter spacings described in Section \ref{Section:Optimization}.



As we can see from either Table \ref{maintable} and Table  \ref{maintable1}, the gradient-search method gives the  same maximum normalized TC as the exhaustive-search method ( $\tau_{opt}=\tau_{opt_{\nabla}}$ ).  However, the gradient-search technique is more efficient because it requires fewer values of $\tau'(\lambda)$ to be evaluated.

\balance

\section{Conclusion} \label{sec_conclusion}
The combination of frequency-hopping, noncoherent CPFSK modulation, and capacity-approaching coding is a sensible choice for modern ad hoc networks.  For such systems, the performance depends critically on the number of frequency-hopping channels, the modulation index, and the code rate.  While these parameters are often chosen arbitrarily, the system performance can be significantly improved by the joint optimization of the three parameters.  The modulation-constrained transmission capacity is an appropriate objective function for the optimization.  Preliminary results using modulation-constrained TC as the objective function suggest that the optimization problem is convex and therefore a good candidate for the gradient-search algorithm proposed in this paper.

The derivation of modulation-constrained transmission capacity required a careful analysis of the outage probability under the assumptions made in this paper.  By extending the analysis, other fading distributions, such as Nakagami, can be considered, as can shadowing.  More sophisticated spatial models can be considered, for instance by imposing a minimum separation among all users.  While more sophisticated models might not be analytically tractable, they are good candidates for the Monte Carlo method proposed in this paper, which requires the random placement of mobiles but does not require the realization of the fading coefficients.

The results presented in this paper are just a sample of what is possible using this methodology.  In addition to considering more sophisticated channel models, future work could consider other network topologies (other than the annular region considered in this paper).  One example of such a network is one where the reference receiver is allowed to move from the center of a disk to its perimeter.  Other types of modulation and reception could be considered, such
as nonbinary CPFSK with multi-symbol reception \cite{valenti:2010}.  Directional antennas could be considered, as could the impact of adjacent-channel interference
due to the effect of spectral splatter.

\bibliographystyle{ieeetr}
\bibliography{VTSymposiumrefs}
\nocite{StocGeo:2009}
\nocite{cardieri:2010}
\nocite{weber:2010}

\balance

\end{document}